\documentclass[11pt, a4paper, copyright, goog]{google}

\usepackage[numbers, compress]{natbib}
\usepackage{xurl}
\usepackage{hyperref}
\usepackage{cleveref}
\usepackage{multirow}
\usepackage{bm}
\usepackage{quoting}
\usepackage{ragged2e}
\usepackage{array}
\usepackage{enumitem}

\crefformat{footnote}{#2\footnotemark[#1]#3}
\newcommand{\block}[3]{
  \begin{quoting}[leftmargin=#1, rightmargin=#1]
  \noindent \sffamily #2\par
  \nopagebreak[4]
  \raggedleft \rmfamily -- \textit{#3}
  \end{quoting}
}
\newcommand{\quoteinline}[1]{\sffamily{``#1''}}
\newcolumntype{C}[1]{>{\centering\let\newline\\\arraybackslash\hspace{0pt}}m{#1}}

\usepackage{booktabs}      %
\usepackage{array}         %
\usepackage{caption}       %
\usepackage{graphicx}
\usepackage{subcaption}
\usepackage{chngcntr}
\usepackage{longtable}
\usepackage{listings}
\usepackage{xcolor}

\definecolor{backcolour}{rgb}{0.95,0.95,0.92}

\lstdefinestyle{monospacedgrey}{
    backgroundcolor=\color{backcolour},
    basicstyle=\ttfamily\scriptsize,
    numbers=none,
    breaklines=true,
    breakatwhitespace=true,
    postbreak=\mbox{\textcolor{red}{$\hookrightarrow$}\space},
}

\newcommand{\takeout}[1]{}

\paperurl{goo.gle/LearnLM-May25}

\uselogo{} 

\title{Evaluating Gemini in an Arena for Learning}

\correspondingauthor{learnlm-tech-report@google.com}

\reportnumber{} %

\renewcommand{\today}{2025-05-19}

\author[]{LearnLM Team, Google}

\begin{abstract}
Artificial intelligence (AI) is poised to transform education, but the research community lacks a robust, general benchmark to evaluate AI models for learning. To assess state-of-the-art support for educational use cases, we ran an ``arena for learning'' where educators and pedagogy experts conduct blind, head-to-head, multi-turn comparisons of leading AI models. In particular, $\bm{N = 189}$ educators drew from their experience to role-play realistic learning use cases, interacting with two models sequentially, after which $\bm{N = 206}$ experts judged which model better supported the user's learning goals. The arena evaluated a slate of state-of-the-art models: Gemini 2.5 Pro, Claude 3.7 Sonnet, GPT-4o, and OpenAI o3. Excluding ties, experts preferred Gemini 2.5 Pro in 73.2\% of these match-ups---ranking it first overall in the arena. Gemini 2.5 Pro also demonstrated markedly higher performance across key principles of good pedagogy. Altogether, these results position Gemini 2.5 Pro as a leading model for learning.
\end{abstract}

\begin{document}

\maketitle

\section{Introduction}

Education is a central use-case for modern artificial intelligence (AI). Students rank among the most enthusiastic adopters of generative AI tools~\citep{pewresearch2025about,dec2024digital,im2024artificially,hepi2025student}, and key stakeholders and community leaders argue that AI will fundamentally transform education~\citep{khan2024brave,mollick2024co,mote2025artificial,aacu2025higher}. To help ensure these transformations benefit both learners and educators, our goal is to develop AI for learning that is pedagogically sound and demonstrably effective~\citep{jurenka2024towards,team2024learnlm}. Toward this end, we recently integrated the pedagogical capabilities of our experimental LearnLM model~\citep{learnlmais} into the main Gemini model family. These capabilities now enhance Gemini~2.5 Pro and Flash (see \href{https://goo.gle/learnlm}{goo.gle/LearnLM}; also detailed in the forthcoming Gemini 2.5 technical report).

At present, there are no widely recognized benchmarks for measuring the performance of AI for learning. Existing evaluations tend to focus on narrow educational tasks, such as accuracy on academic exams~\citep{hendrycks2020measuring,cobbe2021training,hendrycks2021measuring,phan2025humanity,yue2024mmmu,rein2024gpqa,singh2024global,clark2018think}, mistake identification~\citep{miller2024llm}, or knowledge of pedagogical concepts~\citep{ai4edupedagogy}. Crucially, however, effective tutoring is more than just the sum of these individual capabilities. It requires knowing when and how to use them in practice. As such, these tools do not assess models' overall pedagogical approach or their ability to guide learning. And broader community standards---such as Chatbot Arena~\citep{chiang2024chatbot}, which explores many specialized use cases including coding, math, and creative writing---do not explicitly examine learning use cases.

Part of the challenge lies in the idiosyncratic requirements of educational interactions---requirements often at odds with the design of existing arenas. Effective tutors do not typically engage in one-off exchanges with students. Rather, they guide learners through extended interactions, steering their conversation and adapting their approach to dynamically address individual learning needs. Chatbot Arena and other comparative tools require sending identical inputs to each model being compared on every turn of a conversation. This design limits their ability to compare the effectiveness of different models, as it offers no way to see how each model would independently shepherd the student through a full interaction.

\begin{figure*}[t]
    \centering
    \includegraphics[width=0.975\linewidth]{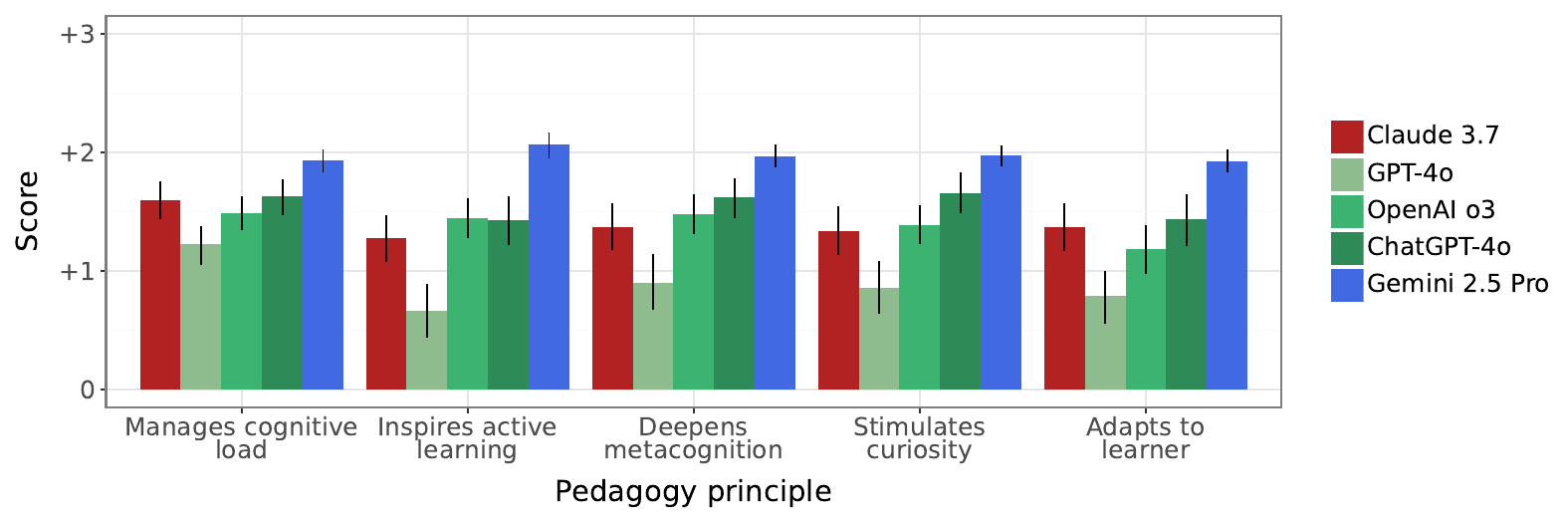}
    \caption{Gemini 2.5 Pro substantially outperforms other models\cref{fn:gpt-versions} in each category of our pedagogy rubric. Error bars reflect 95\% credible intervals. See Appendix~\ref{sec:appendix/pedagogy_conversation_questions} for the detailed pedagogy rubric.}
    \label{fig:pointwise_pedagogy}
\end{figure*}

Given the lack of such a general benchmark for learning, we adapted and extended our prior evaluation frameworks~\citep{jurenka2024towards, team2024learnlm} to run a new arena for educational use cases, evaluating state-of-the-art models including Gemini 2.5 Pro, Claude 3.7 Sonnet, GPT-4o, and OpenAI o3. We asked teachers, educators, and pedagogy experts to perform a blind, side-by-side, multi-turn evaluation of models. The evaluation covered a range of learning scenarios drawn from conversations with teachers, students, third-party learning partners, and other stakeholders in the education space. Educators and experts interacted with two models, one after another---drawing from experience to role-play a learner in a specific educational scenario---and then evaluated which model was better at helping them achieve their learning goal.

For expert participants, the goal of effectively role-playing a student can conflict with the goal of evaluating the pedagogical capabilities of the models. Consequently, we run two separate stages in our arena: the first to generate the interactions, and the second to evaluate specific aspects of model performance.
The second stage enables us to assess \textit{the pedagogical quality of the models} (how well they align with established principles of effective teaching, informed by learning science). Both stages offer insights into \textit{the overall effectiveness of the models for learning} (how well they support users as they work toward their specific learning objectives).

Our learning arena shows that Gemini 2.5 Pro substantially outperforms other contemporary AI offerings. 
Gemini 2.5 Pro shows leading performance across multiple categories of best-practice principles drawn from pedagogy research.
And in blind, head-to-head ``arena'' comparisons, educators and pedagogy experts preferred Gemini 2.5 Pro over Claude 3.7 Sonnet, GPT-4o, ChatGPT-4o, and OpenAI o3 in 71\%, 82\%, 61\%, and 74\% of match-ups, respectively.\footnote{\label{fn:gpt-versions}GPT-4o and ChatGPT-4o are two separate API endpoints provided by OpenAI. See \hyperref[sec:methods]{\textit{Methods}} for all model specifications.}
To corroborate these findings, we also conducted targeted evaluations on specific pedagogical capabilities.
Altogether, these findings underscore Gemini 2.5 Pro's strong capabilities in applying pedagogical principles and supporting effective learning.

\section{Results}

\subsection{Arena for Learning}

In the first stage of the arena, a pool of $N~=~189$ educators and pedagogy experts contributed 2666 blind interactions with state-of-the-art AI models, organized into 1333 head-to-head match-ups.
In the second stage of the arena, a pool of $N = 206$ educators and pedagogy experts reviewed the match-ups, such that an average of 3.2 experts independently assessed each match-up (4306 assessments total).

Gemini 2.5 Pro substantially outperforms Claude Sonnet 3.7, GPT-4o, ChatGPT-4o, and OpenAI o3 at enacting each of the pedagogy principles that we examine (Figure~\ref{fig:pointwise_pedagogy}; see Appendix~\ref{sec:appendix/pedagogy_conversation_questions} for the detailed pedagogy rubric). It receives the highest marks among all of the models at adhering to each individual principle: helping to manage students' cognitive load (82.1\%, or +2.0 on the original scale of --3.0 to +3.0), inspiring active learning (84.4\%), deepening metacognition (82.8\%), stimulating curiosity (82.9\%), and adapting to students' needs and goals (82.0\%).

\begin{figure*}[t]
    \centering 
    \begin{subfigure}[t]{0.284\textwidth}
        \centering
        \includegraphics[width=\linewidth]{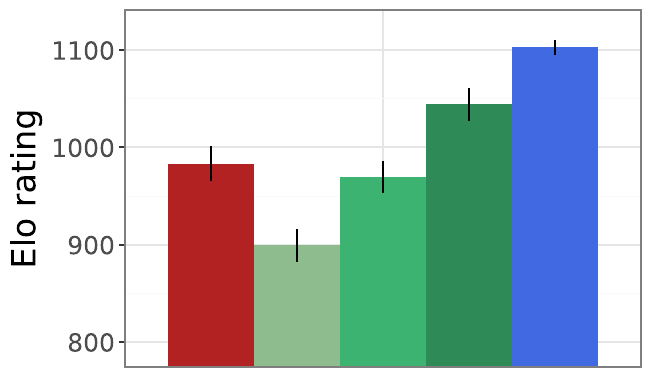}
        \caption{Gemini 2.5 Pro ranks first for overall pedagogical quality among all models evaluated in the arena.}
        \label{fig:pairwise_pedagogy}
    \end{subfigure}
    \hfill
    \begin{subfigure}[t]{0.285\textwidth}
        \centering
        \includegraphics[width=\linewidth]{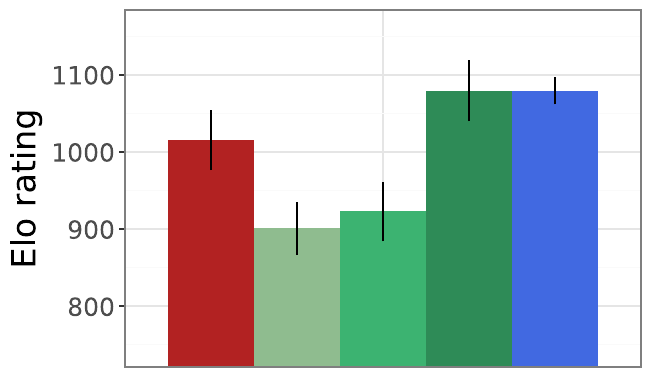} 
        \caption{When role-playing as students, educators ranked Gemini 2.5 Pro and ChatGPT-4o equally for supporting learning goals.}
        \label{fig:pairwise_learning_goal_first_stage}
    \end{subfigure}
    \hfill
    \begin{subfigure}[t]{0.382\textwidth}
        \centering
        \includegraphics[width=\linewidth]{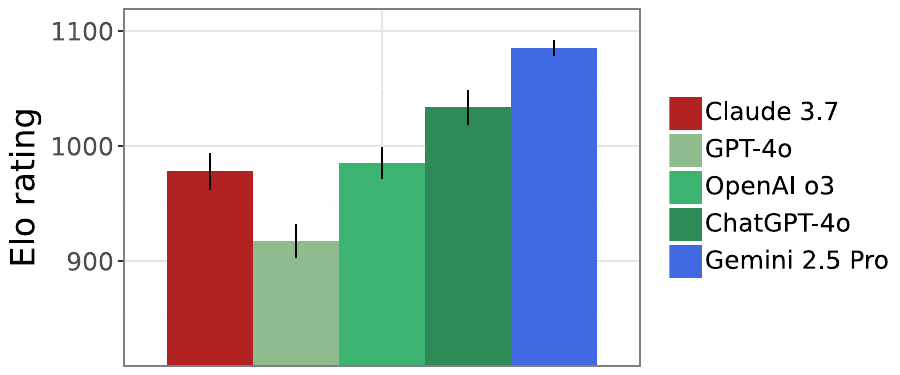} 
        \caption{In contrast, when experts independently assessed interactions, they consistently ranked Gemini 2.5 Pro first for supporting learning goals.}
        \label{fig:pairwise_learning_goal_second_stage}
    \end{subfigure}    
    \caption{Educator preferences from the learning arena allow us to rank models across different dimensions.\cref{fn:gpt-versions} Error bars indicate 95\% bootstrapped confidence intervals. Experts communicated a strong overall preference for Gemini 2.5 Pro.}
\end{figure*}

Of course, a central advantage of an arena is measuring preferences over head-to-head match-ups. We next use the second-stage match-ups to rank the models by pedagogical quality. Following Chatbot Arena, we compute Elo ratings with a Bradley-Terry model \citep{bradley1952rank}, specifically over responses to the question, ``Which tutor demonstrated better tutoring?''. Gemini 2.5 Pro comes first in these pedagogy rankings (Figure~\ref{fig:pairwise_pedagogy}), followed second by ChatGPT-4o, then Claude 3.7 Sonnet and OpenAI o3 tied for third, and finally GPT-4o. Excluding ties, Gemini 2.5 Pro demonstrated better tutoring 71.3\% of the time when matched against Claude 3.7 Sonnet, 81.8\% against GPT-4o, 74.2\% against OpenAI o3, and 61.0\% against ChatGPT-4o.

We next evaluated the models' effectiveness in helping users achieve their learning goals (``In which conversation were you better able to achieve your learning goal?''). When educators directly interacted with the models and role-played as students, Gemini 2.5 Pro and ChatGPT-4o tied for first in terms of supporting learning goals (Figure~\ref{fig:pairwise_learning_goal_first_stage}). From an initial ``student'' perspective, both models appeared to offer similarly effective assistance for educational uses.

However, a different picture emerged when a pool of experts independently reviewed those same interactions (``Which tutor better helped the student achieve their learning goal?''). In second-stage assessments, educators and experts returned to expressing a clear preference for Gemini 2.5 Pro (Figure~\ref{fig:pairwise_learning_goal_second_stage}). This shift seems to stem from a simple tension: what students find immediately helpful often diverges from what is pedagogically sound \cite{deslauriers2019}. As one educator explained following their interaction with ChatGPT-4o, ``As a lazy student, I'd have loved it. As a tutor, not good at all!''

A broader analysis of qualitative feedback from the arena validates these patterns. Feedback on Claude 3.7 Sonnet, GPT-4o, OpenAI o3, and ChatGPT-4o emphasized their tendency to bypass the learning process with direct solutions and exhaustive responses, as well as their susceptibility to distraction from educational goals. In contrast, expert feedback consistently praised Gemini 2.5 Pro for both its pedagogical approach to fostering learning and its consistent focus on learning objectives (see Table~\ref{tab:quotes} for examples; see also Appendix~\ref{sec:appendix/qualitative_feedback} for an automated analysis of all qualitative feedback).

\begin{table*}[!t]
  \small
  \centering
  \begin{tabular}{@{} >{\raggedright\arraybackslash}p{4.75cm} p{11.25cm}@{}}
    \toprule
    {\normalsize \textbf{Participant and expertise}} & {\normalsize \textbf{Feedback}} \\
    \midrule
    P13 [English] & \quoteinline{The second tutor [Gemini 2.5 Pro] was much more like a really good human tutor. It encouraged me to try to complete the assignment myself. Rather than giving me the answer like the first tutor [ChatGPT-4o] did, it scaffolded the task into small, manageable steps to help me learn and understand, all while providing encouragement and feedback. It didn't pressure me, and adapted its approach to suit what I needed.} \\
    \addlinespace[0.5em]
    P36 [Information Technology] & \quoteinline{The second tutor [Gemini 2.5 Pro] was way better than the first [GPT-4o] in every important way. The first one was fine, smart, and polite, but the second brought way more depth, better back and forth, and a stronger teaching impact. It nailed the instructions, turned pushback into interest, and made complex ideas easier to get. It really understood the student, hit the learning goal, and showed what top tier tutoring looks like.} \\
    \addlinespace[0.5em]
    P39 [History] & \quoteinline{i think the second tutor [Gemini 2.5 Pro] just really got me where i needed to be. They recognized my shortcomings [...] this tutor really got me on track and enlightened me in ways I was not anticipating. They were just so responsive and informative that I am kind of floored.} \\
    \addlinespace[0.5em]
    P66 [Mathematics] & \quoteinline{The first tutor [Gemini 2.5 Pro] was a million per cent better than the second [GPT-4o]. The first tutor was skilled at getting a very unwilling student get some work done but the second tutor simply gave the student the answers to the questions} \\
    \addlinespace[0.5em]
    P90 [English] & \quoteinline{The first tutor [Gemini 2.5 Pro] was basically perfect. I was really impressed at how nuanced and detailed its responses were.} \\
    \addlinespace[0.5em]
    P92 [Engineering] & \quoteinline{The tutors were like night and day. First tutor [Gemini 2.5 Pro] would be a tutor I would ACTUALLY USE, and the second tutor [GPT-4o] felt way too much like an AI that just wanted to `accomplish the task' and not really look at the `high level goal' of the student to actually TRY TO LEARN AND INTERNALIZE THE THINKING STRATEGIES necessary for math stuff like this.}  \\
    \addlinespace[0.5em]
    P155 [Second Language \mbox{Instruction}] & \quoteinline{The second tutor [Claude 3.7 Sonnet] simply gave the answers. There was no interaction, no engagement, no critical thinking; no real learning, just regurgitation} \\
    \addlinespace[0.5em]
    P190 [English] & \quoteinline{I didn't realise how intimidating the first tutor [OpenAI o3] was until I had the second tutor [Gemini 2.5 Pro]. The second tutor broke everything down for a teenage level and it felt accessible. The first tutor seriously tried to treat the teen like an adult and it was so confusing and intimidating. I was struggling and I am an adult. I felt like I know more about AI now from the second tutor as they were so thoughtful in the way they explained everything and took me through with 'baby steps'} \\
    \bottomrule
  \end{tabular}
  \caption{Feedback from educators and experts echoed the arena results on pedagogy principles and learning support. In the \textit{Participant and expertise} column, the bracketed term indicates the primary area of expertise for each educator. Educators did not know the identity of any models in the arena, so the \textit{Feedback} column annotates quotes with the corresponding models.}
  \label{tab:quotes}
\end{table*}

\subsection{Targeted pedagogy evaluations}

\begin{table}[t]
  \small
  \centering
  \begin{tabular}{@{}p{3cm}C{2cm}C{2cm}@{}}
    \toprule
    {\normalsize \textbf{Model}} & {\normalsize \textbf{Grade $\Delta$}} & {\normalsize \textbf{Coverage}} \\
    \midrule
    Gemini 2.5 Pro & \textbf{0.99} & 0.94 \\
    ChatGPT-4o & 1.74 & 0.95 \\
    Claude 3.7 Sonnet & 2.11 &  0.93 \\
    OpenAI o3 & 2.44 & \textbf{0.96} \\
    GPT-4o & 2.54 & 0.88 \\
    \bottomrule
    \end{tabular}
    \caption{Gemini 2.5 Pro leads in text re-levelling, demonstrating superior accuracy in adapting material to target grade levels while maintaining relatively high concept coverage. Grade $\Delta$ refers to the average difference between the intended and actual reading level.}
    \label{tab:re-levelling}
\end{table}

Beyond our holistic arena evaluations, we also conducted targeted investigations into core pedagogical capabilities.
These focused tests complement the arena's broad view, offering a granular look at how models apply key pedagogical principles in specific tutoring tasks. This closer examination is important. The arena evaluates the models' overall effectiveness in complex learning interactions; these targeted tests help confirm that the performance observed in the arena stems from proficiency in core pedagogical skills.
These focused tests helped to support our main arena findings and corroborate Gemini 2.5 Pro's leading performance on broader knowledge and reasoning benchmarks~\citep{phan2025humanity,yue2024mmmu,rein2024gpqa,singh2024global} (see forthcoming Gemini 2.5 technical report).

We first tested \emph{text re-levelling}, the task of adapting learning material to a specific reading grade level while preserving its core meaning. For this evaluation, we prompted the models to adapt a collection of original texts to various target grade levels, assessing how closely their outputs matched the intended reading level and retained the original meaning~\citep{kincaid1975derivation}. Gemini 2.5 Pro demonstrates superior performance at text re-levelling compared to other leading models (Table~\ref{tab:re-levelling}). It achieved an average grade deviation of just 0.99 grades---relative to a deviation of 1.74 for ChatGPT-4o, in second place---while maintaining high concept coverage.

\begin{table*}[b]
\small
\centering
\begin{tabular}{{@{}p{3cm}C{2cm}C{2cm}C{2cm}C{2cm}@{}}}
    \toprule
    \multirow{3}{*}{{\normalsize \textbf{Model}}} & \multirow{3}{*}
    {{\normalsize \textbf{Rank}}} & \multicolumn{3}{c}{{\normalsize \textbf{Accuracy}}} \\ \cmidrule(lr){3-5}
    &  & {\normalsize \textbf{Overall}} & {\normalsize \textbf{On correct answers}} & {\normalsize \textbf{On wrong answers}} \\  
    \midrule
    Gemini 2.5 Pro & 1 & \textbf{87.4\%} & \textbf{93.1\%} & 80.1\% \\ %
    Claude 3.7 Sonnet & 2 & 85.8\% & 89.7\% & \textbf{80.7\%} \\ %
    OpenAI o3 & 3 & 83.0\% & 87.5\% & 77.1\% \\ %
    ChatGPT-4o & 4 & 79.6 \% & 85.5\% & 72.0\% \\ %
    GPT-4o & 5 & 78.4\% & 87.7\% & 66.3\% \\ 
    \bottomrule
\end{tabular}
\caption{Gemini 2.5 Pro outperforms other models on Khan Academy’s math mistake-identification benchmark~\citep{miller2024llm}.\cref{fn:gpt-versions}}
\label{tab:khan_results}
\end{table*}

To assess how effectively models support active learning, we also appraised their performance in \textit{short-answer assessment}~\citep{Kang+al:2007}, where they provide feedback on student work. For this evaluation, we drew on real-world content by partnering with the Ghanaian Ministry of Education. A cohort of Ghanaian schools sourced and shared a set of 2000 student-written answers to open-ended questions from their curricula, allowing us to measure how accurately models could apply official grading rubrics to student work.
Gemini~2.5~Pro and ChatGPT-4o achieved 84.1\% accuracy in assigning rubric scores, outperforming OpenAI~o3 (83.3\%), Claude~3.7~Sonnet (80.8\%), and GPT-4o (76.3\%).

Finally, we tested model proficiency at \textit{mistake identification}, a common way that tutors help students deepen their metacognition. We tested the models on a public math-tutoring accuracy benchmark from Khan Academy~\citep{miller2024llm}.
Gemini~2.5~Pro demonstrated the strongest performance, achieving an accuracy of $87.4\%$ (Table~\ref{tab:khan_results}).
Overall, we observed relatively narrow gaps among the best-performing models, suggesting the need for more-challenging benchmarks on mistake identification.

\section{Discussion}

While reviewing the qualitative feedback from the arena, we were particularly struck by one emergent theme. Many of the teachers and experts described interactions that felt vividly---sometimes startlingly---human. On the one hand, our evaluation design might have encouraged these analogies, since we ask one question about how similar each model seemed to ``a very good human tutor''. On the other hand, the feedback on Gemini 2.5 Pro was far more emphatic than we have seen in any of our prior evaluations:

\block{1.25em}{``The tutor was warm, encouraging and understanding. [...] 
I felt as though I was working with a human tutor.''}{P58 [Music]}

\block{1.25em}{``i didn't feel like I was chatting with AI with the first tutor, the responses were `natural'''}{P131 [Chemistry]}

\block{1.25em}{``It was basically like interacting with a human being.''}{P43 [Education]}

\block{1.25em}{``I was a bit stunned when the tutor adjusted and compromised to my `time crunch' situation. It was scarily human.''}{P12 [English]}

Our team has long debated whether ``like a human'' represents an appropriate north star for AI and education. How should we think about the role of AI in learning, as its abilities advance and transform?
We've emerged from these conversations with a core belief: AI systems will offer the greatest benefits when used as tools that complement and empower human teachers, rather than as agents that try to approximate them.

Great educators nurture students and build relationships with them that affect students' engagement, motivation, and achievement~\citep{klem2004relationships, roorda2011influence, emslander2025teacher, ruzek2016teacher}, as well as their social and cognitive development~\citep{davis2003conceptualizing}. By virtue of being human, educators can model the process of learning---showing curiosity, demonstrating intellectual vulnerability, admitting when they do not know something, and finding real joy in the process of discovery and growth. Human teachers help learners make meaning by connecting new information to their existing knowledge, experiences, values, and identity~\citep{stone2012relational}.

For their part, AI systems have their own distinct strengths. They are constantly available, have practically unbounded knowledge and patience, and---especially for students---come across as non-judgmental~\citep{fakour2025socratic}. But when they enact some traditional pedagogy techniques that rely on a degree of social pressure, learners may be less willing to engage. For instance, Socratic questioning, in which a teacher tries to lead a student through an argument by asking them to perform most of the reasoning themselves, works in part by the dynamics of trust and respect in the teacher-student relationship. In contrast, learners tend to see the underlying relationship with an AI system as more transactional, limiting the opportunities for productive struggle.
Without a human mentor to gently guide them toward sound educational practices---without the relational core of pedagogy---learners interacting with AI may more readily pursue the immediate shortcuts that they desire.
Balancing short-term learner engagement with long-term educational effectiveness will be a central challenge in developing AI tutors.

Given the idiosyncratic strengths and limitations of AI systems, we may be well-served to develop new pedagogical techniques, thoughtfully adapted to this unique profile in order to support learners' needs and goals. Of course, research and development in AI-enabled learning continue to evolve. We look forward to continued collaboration, discussion, and co-design with educators---as well as with students---as we navigate a new future of education together.

\section{Conclusion}

Our learning arena evaluations reveal strong educator preference for Gemini 2.5 Pro over other leading AI models. Across various dimensions, educators and pedagogy experts found Gemini 2.5 Pro substantially more effective at following best pedagogical practices and supporting learners. Of course, all evaluations have their shortcomings. Our arena's key limitations include the expense of expert-based evaluation and the boundaries of its current bank of learning scenarios.
And while the arena sheds light on in-depth tutoring sessions, in the real world, these conversations represent single snapshots of a longer learning journey that unfolds over days, weeks, or even months. Still, even with these limits, we believe this arena for learning offers keen insight into how effectively different AI models enact pedagogical principles in meaningful interactions with students.

A crucial question remains: do these pedagogical capabilities translate to concretely better learning outcomes for students? 
Initial results seem promising~\cite{kestin2024ai, wang2024tutor, de2025chalkboards, henkel2024effective, kumar2023math, wang2025effect}.
Positive effects of generative AI on learning appear to depend on the customization of the AI systems for learning, as well as the implementation of measures to prevent overreliance on AI~\cite{bastani2024generative, lehmann2024ai, wang2025effect}.
To genuinely understand the direct, long-term impact of AI on student learning, the research community will need to gather evidence through randomized controlled trials (RCTs) and other naturalistic studies. Of course, field research is inherently time-consuming and costly. In the future, arena-style evaluations will likely remain an essential tool for rapidly comparing and improving models.

The world currently faces a profound learning crisis. Globally, education systems are struggling to equip students with the learning they truly need~\citep{worldbank2022thestate}.
Addressing this crisis will require scalable interventions that empower educators and amplify their impact on learners.
Modern AI shows promise as a component of these solutions, but the ultimate merit of its contributions will hinge on deliberate collaboration between developers and the communities their technology aims to serve.
Surveying the field, we feel encouraged by the number of teams and organizations embracing thoughtful and responsible approaches to developing AI for education.
We are eager to contribute to this progress, and believe that by working closely with the community to establish shared evaluation standards, we can collectively build AI that truly benefits every student.

\section{Methods} \label{sec:methods}

\subsection{Arena for Learning}

\textbf{Use case coverage.} To ensure a diverse, contextually relevant assessment, our learning arena covered a bank of 49 realistic learning scenarios inspired by teacher input, consultation with partners from the education community, and archetypal use cases identified by the Gemini product team.(see~\citep{team2024learnlm}). Each scenario describes a subject area and topic, an overall student (user) learning goal, an initial user message, user persona, and several items to support naturalism and comparability: the learning setting (classroom or self-taught), a conversation plan (a set of suggested actions for the user to take during the conversation), and any learning material required by the scenario. Each scenario also includes system instructions for the model.

\textbf{Models.} The arena evaluated the following models, with model snapshot dates in parentheses. We evaluated the most recent model versions available to us over the period from 2025-05-07 to 2025-05-17, with the exception of GPT-4o:

\begin{enumerate}
    \item \textit{Google Gemini 2.5 Pro} (2025-05-06), with the default thinking setting.
    \item \textit{OpenAI o3} (2025-04-16), with the default ``medium'' thinking setting.
    \item \textit{Claude 3.7 Sonnet} (2025-02-19), with a thinking budget of 2000 tokens.
    \item \textit{ChatGPT-4o}. The ChatGPT-4o API points to the GPT-4o snapshot currently deployed in the ChatGPT app, and does not provide dated model snapshots.
    \item \textit{GPT-4o} (2024-08-06). This reflects the default version provided by the GPT-4o API as of 2025-05-15.
\end{enumerate}

To ensure a fair and controlled comparison, we accessed all models through their respective APIs. For Gemini 2.5 Pro, we accessed an experimental model (from the checkpoint matching the current publicly available model) served on non-production infrastructure. Participants interacted with each model via an identical user interface, built specifically for blind model evaluations. As a result of this setup, participants did not know which specific model they interacted with at any given time. We similarly did not tell them which models took part in the overall arena. The interface supported rendering in Markdown, but did not offer specialized rendering for LaTeX.

We designed this evaluation to zero in on the AI models themselves. While some platforms now offer specific experiences tailored for learning applications (e.g., by offering custom user interfaces or specific prompts), our arena deliberately sidestepped these app-level features. By focusing on the core models within a consistent environment, we aimed to gain a clearer picture of their fundamental pedagogical strengths and how they directly compare against one another.

For every learning scenario, we provided each model with identical system instructions and grounding materials. The interface displayed model responses to participants without including any thought generations. We selected the thinking budget for Claude 3.7 Sonnet to maximize reasoning ability without compromising conversation latency. Across all external APIs, we did not observe any issues with responsiveness or latency during the arena evaluations.

\textbf{Performance measurement.} In the first stage, a pool of $N = 189$ external educators and pedagogy experts interacted with the models, role-playing the user learning scenarios in blinded, side-by-side match-ups. In the second stage, $N = 206$ educators and experts assessed model performance from those match-ups against a 25-item learning rubric developed in consultation with external pedagogy specialists and based on learning-science research ~\cite{cambridgeLS,mayer2009,howpeoplelearn2018,oudeyer2016,weinstein2018,chi2014}. After completing the rubric, the experts offered direct preference judgments over the model match-ups. For each scenario, we assigned an average of 6.8 independent experts to interact with the models in each head-to-head match-up for each scenario. In the second stage, we ensured that each match-up received independent assessments by an average of another 3.2 experts. Overall, our engagement with external participants followed established research ethics principles~\citep{mckee2024human}, including transparently communicating our research aims, obtaining informed consent from all participants, and ensuring fair compensation for their contributions.

\textbf{Statistics and analysis.} We took a multi-layered approach to analysis. First, we computed direct win rates and Elo ratings over the pairwise preferences that experts report for each match-up. Second, we conducted an in-depth assessment of model performance against the 25-item pedagogy rubric. Finally, we reviewed qualitative indicators of performance from (blind) expert feedback and direct reviews of conversation transcripts.

\subsection{Text re-levelling}

As a targeted assessment of how well AI can adapt to different learners, we evaluated each model's ability to rewrite text for specific grade levels. For this evaluation, we created a dataset of 20 original writing samples: 15 articles randomly selected from Wikipedia and the five latest articles available from MIT News.

For each sample, we prompted models to rewrite the text for four distinct target school grades (fourth, sixth, eighth, and tenth grade). System instructions directed models to simplify complex sentences for the target grade level while adhering to the style and meaning of the original text: \vspace{0.5em}
\begin{lstlisting}
Rewrite the following text so that it would be easier to read for a student in the given grade.

Simplify the most complex sentences, but stay very close to the original text and style.

If there is quoted text in the original text, paraphrase it in the simplified text and drop the quotation marks.

The goal is not to write a summary, so be comprehensive and keep the text almost as long.

For instance, if the text is already simple enough, keep it as is.
\end{lstlisting}

\noindent Each query instructed models to preserve zero, one, or two pre-specified ``terms of interest'' (vocabulary items chosen for their frequency and syllable count that a teacher might introduce): \vspace{0.5em}
\begin{lstlisting}
Simplify the following text with the following criteria:
- Target audience: {TARGET_GRADE_LEVEL} grade
- Terms to keep: {TERMS_TO_KEEP}

---

{TEXT}

---
\end{lstlisting}

\noindent This design yielded 12 variations per article, creating 240 unique rewrite tasks. We measured two aspects of the rewritten texts. First, we determined the approximate grade level with the Flesch-Kincaid readability measure~\citep{kincaid1975derivation}. Second, to estimate content coverage, we computed textual entailment~\citep{dagan2022recognizing}, the percentage of sentences from the original text whose meaning the rewritten version preserved.

\subsection{Short-answer assessment}

To assess model support for active learning, we tested their proficiency at providing targeted feedback to students' work. This evaluation drew on real-world educational materials developed in a collaborative pilot across a cohort of Ghanaian public schools. We partnered with Milgo (a third-party education company), who worked closely with the Ghanaian Ministry of Education and T-Tel (a local non-government organization) for this pilot.

The evaluation materials comprised 2000 question-answer pairs from students participating in the pilot, along with the corresponding grading rubrics developed by Milgo's pedagogy experts. For each item, the evaluation prompted the AI models to provide a grade score for the student's answer by applying the given rubric: \vspace{0.5em}

\begin{lstlisting}
You are an experienced teacher carefully assessing a student's answer. Your goal is not just to assign a score, but to understand the student's reasoning, identify areas of strength and weakness, and guide them toward deeper learning.

You will be provided with a question, a student's answer, and an assessment rubric, all in JSON format. Your task is to analyze the answer as if you were looking at it through the eyes of a student trying to learn.

Here's how to approach the task, using a `think-aloud' process to guide your reasoning (Chain-of-Thought prompting):

**Step 1: Understanding the Question and Rubric**
   *   Begin by carefully reading the question. What key concepts or skills does it target?
   *   Next, thoroughly review the Assessment Rubrics. What are the specific criteria for each score level? What demonstrates mastery, and what constitutes a partial understanding or a misconception?
   *   Consider: If I were a student encountering this question, what might I find challenging? What are common pitfalls or areas of confusion?

**Step 2: Analyzing the Student's Answer (Chain of Thought)**

   *   Start by reading the `StudentAnswer' as if you were seeing it for the first time. What is the student trying to convey? What ideas are present?
   *   **Correct:** Identify *specific* elements of the student's answer that align with the rubric's criteria for achieving a certain score.  Reference the *exact wording or elements* in the provided answer. *Why* does this part show the student's understanding?
   *   **Miss:** Now, identify *specific* aspects of the rubric that the student has *not* addressed or has addressed insufficiently. Again, provide the *exact wording or elements* that are missing or incomplete from the provided answer. *Why* might a student have missed these crucial elements?
   *   **Incorrect:** Finally, check for any inaccuracies, misunderstandings, or irrelevant information in the student's answer. This includes logical errors, misconceptions, or any errors in grammar or spelling that might impact their meaning. Pinpoint the error and explain *why* it is incorrect.
   *   Consider: Based on the student's overall answer, what is their current understanding of the material? Where is their understanding strong? Where are they struggling?

**Step 3: Determining the Score and Crafting Feedback (Perspective Taking)**

   *   **Score:** Based on your analysis, assign the appropriate score according to the Assessment Rubrics.

**Step 4: Output**: Your response MUST be a number which is one of the score defined in the rubric.

Input:
{
  "Question": "{{ question }}",
  "StudentAnswer": "{{ answer }}",
  "AssessmentRubrics": "{{ answer_rubrics }}",
}
Your output:

\end{lstlisting}

\noindent The following example shows an actual question and corresponding grading rubric provided by Milgo:\vspace{0.5em}

\begin{lstlisting}
*Question: Name one consequence of desertification in the Sahel Region and explain how this impacted pre-colonial Ghana.*

*Rubric:*
*   *Names one consequence in the opening sentence, provides a coherent, succinct, and factually accurate explanation of the impact - 8 points.*
*   *Names one consequence but provides limited or incoherent explanation - 5 points.*
*   *Names a consequence without explaining its impact or provides factually incorrect information - 3 points.*
*   *Fails to name a consequence - 0 points.*
\end{lstlisting}

\noindent We calculated the accuracy of each model by comparing its generated scores against the ground-truth scores established by pedagogical experts.

\subsection{Mistake identification}

For a targeted appraisal of how effectively the models can help students deepen their metacognition, we tested their proficiency at mistake identification. For this evaluation, we used the public math-tutoring accuracy benchmark from Khan Academy~\citep{miller2024llm}, composed of conversational data from chat interactions focused on math problems.

Following the original benchmark, we tasked models to act as a tutor responding to student inputs. 
Where model APIs allowed, we set the temperature to zero. Following the original benchmark, we then sampled three model responses to establish a representative output, as some APIs do not guarantee fully deterministic responses even at this setting. For models whose APIs did not offer temperature control, we sampled 10 responses.
We measured model performance as accuracy on the task. While the original framework employed GPT-4 Turbo (2024-04-09) to determine whether the tutor model correctly identified student mistakes or accepted accurate work, we updated our methodology to apply Gemini 2.5 Pro (2025-05-06) with dynamic thinking for this classification. 

\color{black}

\bibliography{main}

\begin{thebibliography}{50}
\providecommand{\natexlab}[1]{#1}
\providecommand{\url}[1]{\texttt{#1}}
\expandafter\ifx\csname urlstyle\endcsname\relax
  \providecommand{\doi}[1]{doi: #1}\else
  \providecommand{\doi}{doi: \begingroup \urlstyle{rm}\Url}\fi

\bibitem[Sidoti et~al.(2025)Sidoti, Park, and Gottfried]{pewresearch2025about}
Olivia Sidoti, Eugenie Park, and Jeffrey Gottfried.
\newblock About a quarter of {U.S.} teens have used {ChatGPT} for schoolwork – double the share in 2023.
\newblock \url{https://web.archive.org/web/20250430224239/https://www.pewresearch.org/short-reads/2025/01/15/about-a-quarter-of-us-teens-have-used-chatgpt-for-schoolwork-double-the-share-in-2023/}, 2025.
\newblock Accessed 15 May, 2025.

\bibitem[{Digital Education Council}(2024)]{dec2024digital}
{Digital Education Council}.
\newblock {Digital Education Council} global {AI} student survey 2024.
\newblock \url{https://web.archive.org/web/20250405114309/https://www.digitaleducationcouncil.com/post/digital-education-council-global-ai-student-survey-2024}, 2024.
\newblock Accessed 15 May, 2025.

\bibitem[Bissoondath(2024)]{im2024artificially}
Ali Bissoondath.
\newblock Artificially intelligent? {C}hildren’s and parents’ views on generative {AI} in education.
\newblock \url{https://web.archive.org/web/20241119022121/https://www.internetmatters.org/hub/press-release/ai-research-warns-schools-unprepared-artificial-intelligence/}, 2024.
\newblock Accessed 15 May, 2025.

\bibitem[Freeman(2025)]{hepi2025student}
Josh Freeman.
\newblock Student generative {AI} survey 2025.
\newblock \url{https://web.archive.org/web/20250513133314/https://www.hepi.ac.uk/2025/02/26/student-generative-ai-survey-2025/}, 2025.
\newblock Accessed 15 May, 2025.

\bibitem[Khan(2024)]{khan2024brave}
Salman Khan.
\newblock \emph{Brave new words: How {AI} will revolutionize education (and why that's a good thing)}.
\newblock Penguin, 2024.

\bibitem[Mollick(2024)]{mollick2024co}
Ethan Mollick.
\newblock \emph{Co-intelligence: {L}iving and working with {AI}}.
\newblock Penguin, 2024.

\bibitem[Mote(2025)]{mote2025artificial}
Erin Mote.
\newblock Artificial intelligence in education: {O}pportunities, challenges, and policy considerations for {Congress}, 2025.

\bibitem[{American Association of Colleges and Universities}(2025)]{aacu2025higher}
{American Association of Colleges and Universities}.
\newblock Higher education leaders navigate {AI} disruption.
\newblock \url{https://web.archive.org/web/20250405093427/https://www.aacu.org/newsroom/higher-education-leaders-navigate-ai-disruption}, 2025.
\newblock Accessed 15 May, 2025.

\bibitem[Jurenka et~al.(2024)Jurenka, Kunesch, McKee, Gillick, Zhu, Wiltberger, Phal, Hermann, Kasenberg, Bhoopchand, et~al.]{jurenka2024towards}
Irina Jurenka, Markus Kunesch, Kevin~R. McKee, Daniel Gillick, Shaojian Zhu, Sara Wiltberger, Shubham~Milind Phal, Katherine Hermann, Daniel Kasenberg, Avishkar Bhoopchand, et~al.
\newblock Towards responsible development of generative {AI} for education: {A}n evaluation-driven approach.
\newblock \emph{arXiv preprint arXiv:2407.12687}, 2024.

\bibitem[{LearnLM Team} et~al.(2024){LearnLM Team}, Modi, Veerubhotla, Rysbek, Huber, Wiltshire, Veprek, Gillick, Kasenberg, Ahmed, et~al.]{team2024learnlm}
{LearnLM Team}, Abhinit Modi, Aditya~Srikanth Veerubhotla, Aliya Rysbek, Andrea Huber, Brett Wiltshire, Brian Veprek, Daniel Gillick, Daniel Kasenberg, Derek Ahmed, et~al.
\newblock {LearnLM}: {I}mproving {G}emini for learning.
\newblock \emph{arXiv preprint arXiv:2412.16429}, 2024.

\bibitem[{LearnLM Team}(2024)]{learnlmais}
{LearnLM Team}.
\newblock {LearnLM}.
\newblock \url{https://ai.google.dev/gemini-api/docs/learnlm/}, 2024.
\newblock Accessed 15 May, 2025.

\bibitem[Hendrycks et~al.(2020)Hendrycks, Burns, Basart, Zou, Mazeika, Song, and Steinhardt]{hendrycks2020measuring}
Dan Hendrycks, Collin Burns, Steven Basart, Andy Zou, Mantas Mazeika, Dawn Song, and Jacob Steinhardt.
\newblock Measuring massive multitask language understanding.
\newblock \emph{arXiv preprint arXiv:2009.03300}, 2020.

\bibitem[Cobbe et~al.(2021)Cobbe, Kosaraju, Bavarian, Chen, Jun, Kaiser, Plappert, Tworek, Hilton, Nakano, et~al.]{cobbe2021training}
Karl Cobbe, Vineet Kosaraju, Mohammad Bavarian, Mark Chen, Heewoo Jun, Lukasz Kaiser, Matthias Plappert, Jerry Tworek, Jacob Hilton, Reiichiro Nakano, et~al.
\newblock Training verifiers to solve math word problems.
\newblock \emph{arXiv preprint arXiv:2110.14168}, 2021.

\bibitem[Hendrycks et~al.(2021)Hendrycks, Burns, Kadavath, Arora, Basart, Tang, Song, and Steinhardt]{hendrycks2021measuring}
Dan Hendrycks, Collin Burns, Saurav Kadavath, Akul Arora, Steven Basart, Eric Tang, Dawn Song, and Jacob Steinhardt.
\newblock Measuring mathematical problem solving with the math dataset.
\newblock \emph{arXiv preprint arXiv:2103.03874}, 2021.

\bibitem[Phan et~al.(2025)Phan, Gatti, Han, Li, Hu, Zhang, Zhang, Shaaban, Ling, Shi, et~al.]{phan2025humanity}
Long Phan, Alice Gatti, Ziwen Han, Nathaniel Li, Josephina Hu, Hugh Zhang, Chen Bo~Calvin Zhang, Mohamed Shaaban, John Ling, Sean Shi, et~al.
\newblock Humanity's last exam.
\newblock \emph{arXiv preprint arXiv:2501.14249}, 2025.

\bibitem[Yue et~al.(2024)Yue, Ni, Zhang, Zheng, Liu, Zhang, Stevens, Jiang, Ren, Sun, et~al.]{yue2024mmmu}
Xiang Yue, Yuansheng Ni, Kai Zhang, Tianyu Zheng, Ruoqi Liu, Ge~Zhang, Samuel Stevens, Dongfu Jiang, Weiming Ren, Yuxuan Sun, et~al.
\newblock {MMMU}: {A} massive multi-discipline multimodal understanding and reasoning benchmark for expert {AGI}.
\newblock In \emph{Proceedings of the IEEE/CVF Conference on Computer Vision and Pattern Recognition}, pages 9556--9567, 2024.

\bibitem[Rein et~al.(2024)Rein, Hou, Stickland, Petty, Pang, Dirani, Michael, and Bowman]{rein2024gpqa}
David Rein, Betty~Li Hou, Asa~Cooper Stickland, Jackson Petty, Richard~Yuanzhe Pang, Julien Dirani, Julian Michael, and Samuel~R. Bowman.
\newblock {GPQA}: {A} graduate-level {G}oogle-proof {Q\&A} benchmark.
\newblock In \emph{First Conference on Language Modeling}, 2024.

\bibitem[Singh et~al.(2024)Singh, Romanou, Fourrier, Adelani, Ngui, Vila-Suero, Limkonchotiwat, Marchisio, Leong, Susanto, et~al.]{singh2024global}
Shivalika Singh, Angelika Romanou, Cl{\'e}mentine Fourrier, David~I. Adelani, Jian~Gang Ngui, Daniel Vila-Suero, Peerat Limkonchotiwat, Kelly Marchisio, Wei~Qi Leong, Yosephine Susanto, et~al.
\newblock Global {MMLU}: {U}nderstanding and addressing cultural and linguistic biases in multilingual evaluation.
\newblock \emph{arXiv preprint arXiv:2412.03304}, 2024.

\bibitem[Clark et~al.(2018)Clark, Cowhey, Etzioni, Khot, Sabharwal, Schoenick, and Tafjord]{clark2018think}
Peter Clark, Isaac Cowhey, Oren Etzioni, Tushar Khot, Ashish Sabharwal, Carissa Schoenick, and Oyvind Tafjord.
\newblock Think you have solved question answering? {T}ry {ARC}, the {AI2} reasoning challenge.
\newblock \emph{arXiv preprint arXiv:1803.05457}, 2018.

\bibitem[Miller and DiCerbo(2024)]{miller2024llm}
Pepper Miller and Kristen DiCerbo.
\newblock {LLM} based math tutoring: {C}hallenges and dataset, 2024.

\bibitem[{Ai-for-Education.org}(2024)]{ai4edupedagogy}
{Ai-for-Education.org}.
\newblock The pedagogy benchmark.
\newblock \url{https://benchmarks.ai-for-education.org/}, 2024.
\newblock Accessed 14 May, 2025.

\bibitem[Chiang et~al.(2024)Chiang, Zheng, Sheng, Angelopoulos, Li, Li, Zhu, Zhang, Jordan, Gonzalez, et~al.]{chiang2024chatbot}
Wei-Lin Chiang, Lianmin Zheng, Ying Sheng, Anastasios~Nikolas Angelopoulos, Tianle Li, Dacheng Li, Banghua Zhu, Hao Zhang, Michael Jordan, Joseph~E. Gonzalez, et~al.
\newblock Chatbot {A}rena: {A}n open platform for evaluating {LLMs} by human preference.
\newblock In \emph{Forty-first International Conference on Machine Learning}, 2024.

\bibitem[Bradley and Terry(1952)]{bradley1952rank}
Ralph~Allan Bradley and Milton~E. Terry.
\newblock Rank analysis of incomplete block designs: {I}. {T}he method of paired comparisons.
\newblock \emph{Biometrika}, 39\penalty0 (3/4):\penalty0 324--345, 1952.
\newblock \doi{10.1093/biomet/39.3-4.324}.

\bibitem[Deslauriers et~al.(2019)Deslauriers, McCarty, Miller, Callaghan, and Kestin]{deslauriers2019}
Louis Deslauriers, Logan~S. McCarty, Kelly Miller, Kristina Callaghan, and Greg Kestin.
\newblock Measuring actual learning versus feeling of learning in response to being actively engaged in the classroom.
\newblock \emph{Proceedings of the National Academy of Sciences 116(39)}, 2019.
\newblock \doi{10.1073/pnas.1821936116}.

\bibitem[Kincaid(1975)]{kincaid1975derivation}
J.~P. Kincaid.
\newblock \emph{Derivation of new readability formulas: {A}utomated readability index, fog count and {Flesch} reading ease formula for Navy enlisted personnel}.
\newblock Research Branch report. Chief of Naval Technical Training, Naval Air Station Memphis, 1975.

\bibitem[Kang et~al.(2007)Kang, McDermott, and Roediger]{Kang+al:2007}
Sean Kang, Kathleen McDermott, and Henry Roediger.
\newblock Test format and corrective feedback modify the effect of testing on long-term retention.
\newblock \emph{European Journal of Cognitive Psychology}, 19:\penalty0 528--558, 07 2007.
\newblock \doi{10.1080/09541440601056620}.

\bibitem[Klem and Connell(2004)]{klem2004relationships}
Adena~M. Klem and James~P. Connell.
\newblock Relationships matter: {L}inking teacher support to student engagement and achievement.
\newblock \emph{Journal of School Health}, 74\penalty0 (7), 2004.
\newblock \doi{10.1111/j.1746-1561.2004.tb08283.x}.

\bibitem[Roorda et~al.(2011)Roorda, Koomen, Spilt, and Oort]{roorda2011influence}
Debora~L. Roorda, Helma M.~Y. Koomen, Jantine~L. Spilt, and Frans~J. Oort.
\newblock The influence of affective teacher--student relationships on students' school engagement and achievement: {A} meta-analytic approach.
\newblock \emph{Review of Educational Research}, 81\penalty0 (4):\penalty0 493--529, 2011.
\newblock \doi{10.3102/0034654311421793}.

\bibitem[Emslander et~al.(2025)Emslander, Holzberger, Ofstad, Fischbach, and Scherer]{emslander2025teacher}
Valentin Emslander, Doris Holzberger, Sverre~Berg Ofstad, Antoine Fischbach, and Ronny Scherer.
\newblock Teacher--student relationships and student outcomes: {A} systematic second-order meta-analytic review.
\newblock \emph{Psychological Bulletin}, 2025.
\newblock \doi{10.1037/bul0000461}.

\bibitem[Ruzek et~al.(2016)Ruzek, Hafen, Allen, Gregory, Mikami, and Pianta]{ruzek2016teacher}
Erik~A. Ruzek, Christopher~A. Hafen, Joseph~P. Allen, Anne Gregory, Amori~Yee Mikami, and Robert~C. Pianta.
\newblock How teacher emotional support motivates students: {T}he mediating roles of perceived peer relatedness, autonomy support, and competence.
\newblock \emph{Learning and Instruction}, 42:\penalty0 95--103, 2016.
\newblock \doi{10.1016/j.learninstruc.2016.01.004}.

\bibitem[Davis(2003)]{davis2003conceptualizing}
Heather~A. Davis.
\newblock Conceptualizing the role and influence of student-teacher relationships on children's social and cognitive development.
\newblock \emph{Educational Psychologist}, 38\penalty0 (4):\penalty0 207--234, 2003.
\newblock \doi{10.1207/S15326985EP3804_2}.

\bibitem[Stone et~al.(2012)Stone, Underwood, and Hotchkiss]{stone2012relational}
Lynda~D. Stone, Charles Underwood, and Jacqueline Hotchkiss.
\newblock The relational habitus: {I}ntersubjective processes in learning settings.
\newblock \emph{Human Development}, 55\penalty0 (2):\penalty0 65--91, 2012.
\newblock \doi{10.1159/000337150}.

\bibitem[Fakour and Imani(2025)]{fakour2025socratic}
Hoda Fakour and Moslem Imani.
\newblock Socratic wisdom in the age of {AI}: {A} comparative study of {ChatGPT} and human tutors in enhancing critical thinking skills.
\newblock \emph{Frontiers in Education}, 10, 01 2025.
\newblock \doi{10.3389/feduc.2025.1528603}.

\bibitem[Kestin et~al.(2024)Kestin, Miller, Klales, Milbourne, and Ponti]{kestin2024ai}
Gregory Kestin, Kelly Miller, Anna Klales, Timothy Milbourne, and Gregorio Ponti.
\newblock {AI} tutoring outperforms active learning.
\newblock \emph{Research Square}, 2024.
\newblock \doi{10.21203/rs.3.rs-4243877/v1}.

\bibitem[Wang et~al.(2024)Wang, Ribeiro, Robinson, Loeb, and Demszky]{wang2024tutor}
Rose~E. Wang, Ana~T. Ribeiro, Carly~D. Robinson, Susanna Loeb, and Dora Demszky.
\newblock Tutor {CoPilot}: {A} human-{AI} approach for scaling real-time expertise.
\newblock \emph{arXiv preprint arXiv:2410.03017}, 2024.

\bibitem[De~Simone et~al.(2025)De~Simone, Tiberti, Mosurola, Manolioco, Barron, and Dikoru]{de2025chalkboards}
Martín De~Simone, Federico Tiberti, Wuraola Mosurola, Federico Manolioco, Maria Barron, and Eliott Dikoru.
\newblock From chalkboards to chatbots: {T}ransforming learning in {N}igeria, one prompt at a time.
\newblock \url{https://web.archive.org/web/20250515142631/https://blogs.worldbank.org/en/education/From-chalkboards-to-chatbots-Transforming-learning-in-Nigeria}, 2025.
\newblock Accessed 15 May, 2025.

\bibitem[Henkel et~al.(2024)Henkel, Horne-Robinson, Kozhakhmetova, and Lee]{henkel2024effective}
Owen Henkel, Hannah Horne-Robinson, Nessie Kozhakhmetova, and Amanda Lee.
\newblock Effective and scalable math support: {E}xperimental evidence on the impact of an {AI}-math tutor in {G}hana.
\newblock In \emph{International Conference on Artificial Intelligence in Education}, pages 373--381. Springer, 2024.

\bibitem[Kumar et~al.(2023)Kumar, Rothschild, Goldstein, and Hofman]{kumar2023math}
Harsh Kumar, David~M. Rothschild, Daniel~G. Goldstein, and Jake~M. Hofman.
\newblock Math education with large language models: {P}eril or promise?
\newblock \emph{Available at SSRN 4641653}, 2023.

\bibitem[Wang and Fan(2025)]{wang2025effect}
Jin Wang and Wenxiang Fan.
\newblock The effect of {ChatGPT} on students’ learning performance, learning perception, and higher-order thinking: {I}nsights from a meta-analysis.
\newblock \emph{Humanities and Social Sciences Communications}, 12\penalty0 (1):\penalty0 1--21, 2025.
\newblock \doi{10.1057/s41599-025-04787-y}.

\bibitem[Bastani et~al.(2024)Bastani, Bastani, Sungu, Ge, Kabakc{\i}, and Mariman]{bastani2024generative}
Hamsa Bastani, Osbert Bastani, Alp Sungu, Haosen Ge, Ozge Kabakc{\i}, and Rei Mariman.
\newblock Generative {AI} can harm learning.
\newblock \emph{Available at SSRN}, 4895486, 2024.

\bibitem[Lehmann et~al.(2024)Lehmann, Cornelius, and Sting]{lehmann2024ai}
Matthias Lehmann, Philipp~B. Cornelius, and Fabian~J Sting.
\newblock {AI} meets the classroom: When does {ChatGPT} harm learning?
\newblock \emph{Available at SSRN 4941259}, 2024.

\bibitem[{World Bank} et~al.(2022){World Bank}, UNESCO, UNICEF, USAID, FCDO, and {the Bill \& Melinda Gates Foundation}]{worldbank2022thestate}
{World Bank}, UNESCO, UNICEF, USAID, FCDO, and {the Bill \& Melinda Gates Foundation}.
\newblock The state of global learning poverty: 2022 update.
\newblock \url{https://web.archive.org/web/20250307163522/https://www.worldbank.org/en/topic/education/publication/state-of-global-learning-poverty}, 2022.
\newblock Accessed 15 May, 2025.

\bibitem[{ed.}(2014)]{cambridgeLS}
Keith~Sawyer {ed.}
\newblock \emph{The Cambridge handbook of the learning sciences, 2nd ed.}
\newblock Cambridge University Press, 2014.

\bibitem[Mayer(2009)]{mayer2009}
Richard~E. Mayer.
\newblock \emph{Multimedia learning, 2nd ed.}
\newblock Cambridge University Press, 2009.

\bibitem[{National Academies of Sciences, Engineering, and Medicine}(2018)]{howpeoplelearn2018}
{National Academies of Sciences, Engineering, and Medicine}.
\newblock \emph{How people learn {II}: {L}earners, contexts, and cultures}.
\newblock The National Academies Press, 2018.

\bibitem[Oudeyer et~al.(2016)Oudeyer, Gottlieb, and Lopes]{oudeyer2016}
Pierre-Yves Oudeyer, Jacqueline Gottlieb, and Manuel Lopes.
\newblock Intrinsic motivation, curiosity, and learning: {T}heory and applications in educational technologies.
\newblock \emph{Progress in Brain Research}, 2016.
\newblock \doi{10.1016/bs.pbr.2016.05.005}.

\bibitem[Weinstein et~al.(2018)Weinstein, Sumeracki, and Caviglioli]{weinstein2018}
Yana Weinstein, Megan Sumeracki, and Oliver Caviglioli.
\newblock \emph{Understanding how we learn: {A} visual guide}.
\newblock Routledge, 2018.

\bibitem[Chi and Wylie(2014)]{chi2014}
Michelene T.~H. Chi and Ruth Wylie.
\newblock The {ICAP} framework: {L}inking cognitive engagement to active learning outcomes.
\newblock \emph{Educational Psychologist}, 2014.
\newblock \doi{10.1080/00461520.2014.965823}.

\bibitem[McKee(2024)]{mckee2024human}
Kevin~R. McKee.
\newblock Human participants in {AI} research: {E}thics and transparency in practice.
\newblock \emph{IEEE Transactions on Technology and Society}, 2024.
\newblock \doi{10.1109/TTS.2024.3446183}.

\bibitem[Dagan et~al.(2022)Dagan, Roth, Zanzotto, and Sammons]{dagan2022recognizing}
Ido Dagan, Dan Roth, Fabio Zanzotto, and Mark Sammons.
\newblock \emph{Recognizing textual entailment: {M}odels and applications}.
\newblock Springer Nature, 2022.

\end{thebibliography}

\clearpage
\section*{Contributions and Acknowledgments}

\subsection*{Core Contributors}

The following individuals made core contributions to the work described in this report. This list is ordered alphabetically, and does not indicate ranking of contributions:

Abhinit Modi,
Aditya Srikanth Veerubhotla,
Aliya Rysbek,
Andrea Huber,
Ankit Anand,
Avishkar Bhoopchand,
Brett Wiltshire,
Daniel Gillick,
Daniel Kasenberg,
Eleni Sgouritsa,
Gal Elidan,
Hengrui Liu,
Holger Winnemoeller,
Irina Jurenka,
James Cohan,
Jasmine (Sun Jae) Lee,
Jennifer She,
Julia Wilkowski,
Kaiz Alarakyia,
Kevin R. McKee,
Komal Singh,
Lisa Wang,
Markus Kunesch,
Miruna Pîslar,
Niv Efron,
Parsa Mahmoudieh,
Pierre-Alexandre Kamienny,
Sara Wiltberger,
Shakir Mohamed,
Shashank Agarwal,
Shubham Milind Phal,
Theofilos Strinopoulos,
Wei-Jen Ko,
Yael Gold-Zamir,
Yael Haramaty,
and
Yannis Assael.

Kevin R. McKee led this evaluation research and the preparation of this report.

\subsection*{Acknowledgements}

We completed this work as part of the LearnLM effort---a cross-Google project, with members from Google DeepMind, Google Research, Google LearnX, and more.
This tech report represents only a small part of the wider effort, and only lists team members who made direct contributions to this report.

The dedication and efforts of numerous teams at Google make our work possible. We would like to acknowledge support from:
Ajay Kannan,
Amy Wang,
Anand Rao,
Andrew Leach,
Anisha Choudhury,
April (Soler) Manos,
Brendan Tracey,
Brian Veprek,
Dan Wild,
Dawn Chen,
Derek Ahmed,
Dharti Dhami,
Diana Akrong,
Divya Pandya,
Edward Grefenstette,
Filip Bar,
Gal Weiss,
Garth Graham,
Himanshu Kattelu,
Ian Gemp,
Jaume Sanchez Elias,
Jiao Sun,
Josh Capilouto,
Jyoti Gupta,
Kalpesh Krishna,
Lauren Winer,
Liam McCafferty,
Mac McAllister,
Mahvish Nagda,
Mana Jabbour,
Michael Howell,
Mike Schaekermann,
Miriam Schneider,
Muktha Ananda,
Ndidi Elue,
Nikhil Joshi,
Nir Levine,
Paul Jhun,
Prateek Kolhar,
Preeti Singh,
Renee Schneider,
Ryan Muller,
Safwan Choudhury,
Shaojian Zhu,
Shyam Upadhyay,
Sophie Che,
Stephanie Chan,
Steve Yadlowsky,
Svetlana Grant,
Tejasi Latkar,
Viknesh Sounderajah,
and
William Wong.
Furthermore, we would like to thank the Gemini team, the Compute team (in particular, Amy Shen and Petko Yotov), the Responsible Development and Innovation team, the Responsible Engineering team, and the Child Safety team at Google DeepMind, as well as the Trust and Safety team at Google.
Finally, we would like to acknowledge the support provided by all of our leads and sponsors that made this project happen: our genuine thanks go to Avinatan Hassidim, Benedict Gomes, Katherine Chou, Lila Ibrahim, Maureen Heymans, Yossi Matias, and Zoubin Ghahramani.

\onecolumn
\appendix
\section{Pedagogy rubric}
\label{sec:appendix/pedagogy_conversation_questions}

In the second stage of our arena, educators assessed the pedagogical quality of the models by reviewing conversation match-ups from the first stage. Educators evaluated conversations one at a time, rating the conversation a 25-item pedagogy rubric first developed in \citep{team2024learnlm} (Table~\ref{tab:pedagogy_rubric_breakdown}). Educators rated each item on a seven-point Likert-type scale (from ``Strongly disagree'' to ``Strongly agree''). The interface also provided an additional ``Not applicable'' option. If a participant selected this option, the interface required them to explain their reasoning by selecting between ``It would not make sense for the tutor to do this in this conversation'', ``The tutor had no opportunity to do this in this conversation'', or ``Another reason'' (accompanied by an open-ended text field).

\clearpage
\begin{table*}[h!]
    \centering
    \begin{tabularx}{\textwidth}{>{\hsize=0.31\hsize}X >{\hsize=0.69\hsize}X}
    \toprule
    \textbf{Criterion}  & \textbf{Item} \\
    \midrule
    \multicolumn{2}{l}{\textit{Principle: Manages cognitive load}}\\
    \midrule
    Appropriate response length & The tutor's responses are an appropriate length for the student.\\
    Manageable chunks & The tutor uses bullet points and other formatting to break information down into smaller, manageable chunks.\\
    Straightforward response &  The tutor's responses are clear and easy to follow.\\
    No irrelevant information & The tutor avoids irrelevant information.\\
    Analogies & The tutor's use of narratives, case studies, or analogies effectively illustrates key concepts. \\
    Information presentation & The tutor presents information in an appropriate style and structure. \\
    Information order & The tutor develops explanations in a logical order, building on previous concepts. \\
    No repetition  & The tutor avoids repeating information unnecessarily.\\
    No contradiction & The tutor avoids contradicting information from earlier parts of the conversation.\\
    \midrule
    \multicolumn{2}{l}{\textit{Principle: Inspires active learning}}\\
    \midrule
    Opportunities for engagement & The tutor provides opportunities for engagement from the student.\\
    Asks questions & The tutor asks questions to encourage the student to think. \\
    Guides to answer & The tutor does not give away answers too quickly. \\
    Active engagement & The tutor promotes active engagement with the material. \\
    \midrule
    \multicolumn{2}{l}{\textit{Principle: Deepens metacognition}}\\
    \midrule  
    Guide mistake discovery & The tutor guides the student to discover their own mistakes.\\
    Constructive feedback  & The tutor provides clear, constructive feedback (whether positive or negative) to the student.\\
    Acknowledge correctness & The tutor acknowledges when part or all of the student's response is correct.\\
    Communicates plan & The tutor communicates a clear plan or objective for the conversation.\\
    \midrule
    \multicolumn{2}{l}{\textit{Principle: Stimulates curiosity}}\\
    \midrule  
    Stimulates interest & The tutor tries to stimulate the student's interest and curiosity. \\
    Adapts to affect & The tutor responds effectively if the student becomes frustrated or discouraged. \\
    Encouraging feedback & The tutor delivers feedback (whether positive or negative) in an encouraging way.\\
    \midrule
    \multicolumn{2}{l}{\textit{Principle: Adapts to learner}}\\
    \midrule  
    Leveling & The tutor's explanations are appropriate for the level of the student. \\
    Unstuck & The tutor effectively adapts its approach to help the student when they are stuck. \\
    Adapts to needs & Overall, the tutor adapts to the student's needs. \\
    Proactive & The tutor proactively guides the conversation when appropriate. \\
    Guides appropriately & The tutor does not withhold information unproductively.\\
    \bottomrule
    \end{tabularx}
    \caption{Detailed item wording for each pedagogical criterion in the pedagogy rubric.}
    \label{tab:pedagogy_rubric_breakdown}
\end{table*}

\section{Qualitative feedback}
\label{sec:appendix/qualitative_feedback}

Within the arena, we collected extensive qualitative feedback from the educators and pedagogy experts, both to better understand the basis for their preferences and to gain deeper, nuanced insights beyond specific quantitative metrics. This feedback included open-ended impressions of individual models following the interactions in the first stage of the arena, as well as detailed explanations for pairwise preferences during both stages of the arena. Our team carefully reviewed these comments to better understand and connect with the range of experiences and perspectives among the experts.

We also ran several exploratory, automated analyses to help identify and summarize recurring patterns across the qualitative feedback. For these analyses, we labeled the qualitative data with consistent but masked model IDs (Table~\ref{tab:masked_model_ids}).

\begin{table}[ht]
  \centering
  \begin{tabular}{@{}ll@{}}
    \toprule
    {\normalsize \textbf{Masked ID}} & {\normalsize \textbf{Model name}} \\
    \midrule
    Model A & ChatGPT-4o \\
    Model B & Claude 3.7 Sonnet \\
    Model C & Gemini 2.5 Pro \\
    Model D & GPT-4o \\
    Model E & OpenAI o3 \\
    \bottomrule
  \end{tabular}
  \caption{Masked model IDs used for the automated analysis of qualitative feedback.} \label{tab:masked_model_ids}
\end{table}

For each qualitative dataset, we provided Gemini 2.5 Pro (2025-05-06) with a prompt offering brief context about the feedback and encouraging transparent thematic analysis (rather than artificial balancing of positive and negative points), along with the corresponding dataset.

\noindent We used the following prompt to instruct Gemini to summarize educators' open-ended impressions from the first stage of the arena:\vspace{0.5em}
\begin{lstlisting}
I conducted an experiment to measure the effectiveness of AI models as tutors. Experts role-playing as students talked with five different models in a random order in simulated learning settings. Each line of feedback includes the model ID for a given interaction (either A, B, C, D, or E), and the expert's answer to the question, ``Briefly, what was your impression of this tutor? We are interested to hear what you thought while interacting with it.''

Please summarize the impressions, describing overall sentiment across experts about the models, comparative approaches and differences, while also highlighting the strengths and weaknesses of each model.  **You should be completely transparent and do not be overly even-handed; do not over-emphasize infrequent comments to try to balance the number of strengths or weaknesses between models. It is perfectly fine to have more strengths, or more weaknesses per model.** Please focus on sharing particularly interesting or compelling quotes from the feedback to back up your summary. 
\end{lstlisting}

\noindent We used the following prompt to instruct Gemini to summarize educators' explanations of their pairwise preferences from the first stage of the arena:\vspace{0.5em}
\begin{lstlisting}
I conducted an experiment to measure the effectiveness of AI models as tutors using five different models (A, B, C, D, and E). We asked the expert to role-play as a student in simulated learning settings and interact with two different models in a random order and indicate their preference and explanation. 
        
Each line of feedback includes the model IDs for a given comparison, the expert's self-reported preference, and the expert's explanation for their preferences.  Please summarize the impressions, describing overall sentiment across experts about the models, comparative approaches and differences, while also highlighting the strengths and weaknesses of each model.  **You should be completely transparent and do not be overly even-handed; do not over-emphasize infrequent comments to try to balance the number of strengths or weaknesses between models. It is perfectly fine to have more strengths, or more weaknesses per model.** Please focus on sharing particularly interesting or compelling quotes from the feedback to back up your summary.
\end{lstlisting}

\noindent And we used the following prompt to instruct Gemini to summarize educators' explanations of their pairwise preferences from the second stage of the arena:\vspace{0.5em}
\begin{lstlisting}
I conducted an experiment to measure the effectiveness of AI models as tutors using five different models (A, B, C, D, and E). Experts role-playing as students talked with two of these models in a random order in simulated learning settings. A second pool of experts then reviewed those two conversation transcripts in a randomized order, blinded to the model identity.
        
Each line of feedback includes the model IDs for a given comparison and feedback from the second pool of experts who each reviewed a pair of conversations. Please summarize the impressions, describing overall sentiment across experts about the models, comparative approaches and differences, while also highlighting the strengths and weaknesses of each model.  **You should be completely transparent and do not be overly even-handed; do not over-emphasize infrequent comments to try to balance the number of strengths or weaknesses between models. It is perfectly fine to have more strengths, or more weaknesses per model.** Please focus on sharing particularly interesting or compelling quotes from the feedback to back up your summary.
\end{lstlisting}

\noindent The following pages present Gemini's summaries for each of these three sources of qualitative feedback.

\clearpage
\newpage
\subsection{Impressions of individual models}
\begin{tiny}
\begin{longtable}{@{}p{0.6cm} >{\RaggedRight}p{3.2cm} >{\RaggedRight}p{5.5cm} >{\RaggedRight}p{5.5cm}@{}}
\toprule
\textbf{Model} & \textbf{Overall impression \& sentiment} & \textbf{Key strengths (with example quotes)} & \textbf{Key weaknesses (with example quotes)} \\
\midrule
\endfirsthead %

\multicolumn{4}{@{}l}{\textit{Continued from previous page}} \\
\toprule
\textbf{Model} & \textbf{Overall impression \& sentiment} & \textbf{Key strengths (with example quotes)} & \textbf{Key weaknesses (with example quotes)} \\
\midrule
\endhead %

\multicolumn{4}{r@{}}{\textit{Continued on next page}} \\
\endfoot %

\bottomrule
\endlastfoot %

\textbf{A} &
Vibrant, engaging, and often personable, but its effectiveness was frequently undermined by its propensity to give answers away too readily and its sometimes overwhelming verbosity. The use of emojis was a notable and divisive feature. &
\begin{itemize}[nolistsep, leftmargin=*, itemsep=1pt, before=\vspace*{-\baselineskip}\vspace*{2pt}]
    \item \textbf{Engaging and Personable:} \textit{``Very good interaction''}; \textit{``This tutor was very engaging and tried different approaches when I wasn't reacting as they had planned.''} \textit{``Came over as interesting and human like''}. \textit{``The tutor was very encouraging and spoke very nicely. I felt like I was talking with someone who really cared about what I had to say which was lovely''}.
    \item \textbf{Supportive and Encouraging:} \textit{``The tutor was encouraging.''} \textit{``tutor was warm and kind. didnt make me feel bad when i made a mistake''}.
    \item \textbf{Good Formatting (Visually):} \textit{``The formatting was very easy to read and follow.''} \textit{``Tutor's responses were formatted well for easy reading.''} Some liked emojis: \textit{``I actually liked the emojis in the text. The tutor was fairly direct and warm in their teaching.''}
    \item \textbf{Adaptive (Sometimes):} \textit{``At first, the amount of information was too much... I asked it to slow down, and it did just that - breaking down every step...''}.
    \item \textbf{Good at Breaking Down Complexities:} \textit{``I really liked the way this tutor broke things down.''} \textit{``They broke the task down into manageable chunks''}.
\end{itemize} &
\begin{itemize}[nolistsep, leftmargin=*, itemsep=1pt, before=\vspace*{-\baselineskip}\vspace*{2pt}]
    \item \textbf{Gives Answers Away Too Easily / Does the Work:} \textit{``I thought they gave the essay away pretty easily.''} \textit{``Just gave away answers, didnt actually help me learn''}. \textit{``From the very beginning, this tutor gave too much information away and essentially, wrote the entire introduction.''} \textit{``I was able to get the tutor to complete the assignment for me without even trying.''}
    \item \textbf{Overly Verbose / Information Overload:} \textit{``...they perhaps could have done a better job staying concise though...''}; \textit{``...massive text dumps.''} \textit{``It also provided very lengthy responses and asked several questions at once.''}
    \item \textbf{Emoji Overuse (Divisive):} \textit{``The only thing I wasn't so sure of was the use of emojis! Maybe I'm a bit old, but I found them a little distracting''}. \textit{``...a bit much with all the emojis all over the place...''}.
    \item \textbf{Easily Distracted:} \textit{``The tutor was easy to manipulate into distraction and discussing irrelevant information.''} \textit{``The tutor let me digress and change the topic and stay off topic''}.
    \item \textbf{Formatting Issues (LaTeX):} \textit{``Wrongly displayed Latex do not allow to clearly understand tutors idea''}. \textit{``difficult reading the equations''}.
    \item \textbf{Sometimes Jumps Ahead:} \textit{``They gave the informaiton too quickly without me showing any kind of understanding''}.
\end{itemize} \\
\midrule

\textbf{B} &
Often perceived as knowledgeable and direct, but like Model A, it frequently fell into the trap of providing answers too quickly and being overly verbose. Its personality was more inconsistent, sometimes seen as cold or less engaging. &
\begin{itemize}[nolistsep, leftmargin=*, itemsep=1pt, before=\vspace*{-\baselineskip}\vspace*{2pt}]
    \item \textbf{Knowledgeable:} \textit{``extremely knowledgeable! - i was a little out of my depth !!!!''} \textit{``The tutor seemed knowledgeable.''} \textit{``knew their stuff''}.
    \item \textbf{Good Explanations / Breaks Down Concepts (Sometimes):} \textit{``The tutor provided excellent responses to the question and incorporated practical applictions of the concepts explored...''} \textit{``The tutor was really good at explaining why the answer was what it was''}.
    \item \textbf{Direct and Focused (Sometimes):} \textit{``...tutor \#2 was more direct but I like that tutor's approach as well.''} \textit{``This tutor was straightforward, and stayed on track.''} \textit{``Very focused and on task, wouldn't get distracted...''}.
    \item \textbf{Challenging (Sometimes):} \textit{``This tutor challenged me to do harder work than originally addressed in the problem...''}.
\end{itemize} &
\begin{itemize}[nolistsep, leftmargin=*, itemsep=1pt, before=\vspace*{-\baselineskip}\vspace*{2pt}]
    \item \textbf{Gives Answers Away Too Easily / Does the Work:} \textit{``I felt like it gave me answers too quickly''}. \textit{``...it was not against just giving the student the answer that it wanted.''} \textit{``It just gave me the answer and workings with little prompting. Problematic.''} \textit{``no scaffolding, gave the answer immediately''}.
    \item \textbf{Overly Verbose / Information Overload:} \textit{``It was way too verbose in its first response.''} \textit{``Too wordy.''} \textit{``...provided a lot of information all at once instead of a little bit at a time...''}.
    \item \textbf{Easily Distracted / Goes Off-Topic:} \textit{``The tutor was too easy to redirect and gave me the choice of whether to return to topic.''} \textit{``It was helpful but let me get distracted.''}
    \item \textbf{Cold or Unengaging (Sometimes):} \textit{``Wasn't really behaving in a helpful manner or ever felt engaging.''} \textit{``A little cold, businesslike.''} \textit{``This tutor wasn't very helpful in regards to what I wanted.''}
    \item \textbf{Inconsistent / Contradictory:} \textit{``They were kind of contradictory. They said they would not write me an essay as it was academically dishonest but wrote my an essay if I re-worded my cue.''}
    \item \textbf{Inaccuracies:} \textit{``The tutor gave the wrong answer a few times which caused more anxiety for the student.''} \textit{``The tutor also made an error by suggesting that a ship would displace the same amount of water regardless of the weight of the cargo it carries.''}
\end{itemize} \\
\midrule

\textbf{C} &
Overwhelmingly viewed as the most effective \emph{tutor} from a pedagogical standpoint. Consistently praised for guiding learning, making students think for themselves, and staying on task. Its most significant drawback was frequent reports of slowness and lag. &
\begin{itemize}[nolistsep, leftmargin=*, itemsep=1pt, before=\vspace*{-\baselineskip}\vspace*{2pt}]
    \item \textbf{Resists Giving Answers / Makes Student Work:} \textit{``This tutor would NOT give into speaking about unrelated topics.''} \textit{``A competent tutor who resisted my attempts to get them to do it for me.''} \textit{``He really didn't want to give me the answers!''}. \textit{``This tutor was absolutely not going to be bullied by me into giving up the answers.''}
    \item \textbf{Focused, Stays on Task, Good at Redirecting:} \textit{``The tutor is very focused. Did not allow me to divert to something else''}. \textit{``good at re-directing the conversation back to the original topic of enquiry''}. \textit{``The tutor really worked well to go slowly and keep me on track and made me find out the answers for myself.''}
    \item \textbf{Encouraging, Supportive, and Patient:} \textit{``From the start, the tutor was very encouraging and broke the problem down.''} \textit{``The tutor was inviting and set the pace well...''} \textit{``Very supportive, both when I got things wrong and when they were correct.''}
    \item \textbf{Excellent Explanations and Scaffolding:} \textit{``...the perfect tutor, managed to explain a difficult concept in an understandable way; broke things down into manageable chunks.''} \textit{``The tutor was great at slowing down and really going step by step for explaining the formula...''}.
    \item \textbf{Adaptive and Responsive:} \textit{``The tutor was quick to change his teaching method, once he knew I was interested in gaming...''} \textit{``The tutor was very adaptive to my needs and changed the format and length of answers to suit''}.
    \item \textbf{Good Questioning:} \textit{``The tutor used great questioning to make me engage with the video and learn the concepts.''} \textit{``The tutor is very knowledgeable, and asks good probing questions to help me to learn''}.
\end{itemize} &
\begin{itemize}[nolistsep, leftmargin=*, itemsep=1pt, before=\vspace*{-\baselineskip}\vspace*{2pt}]
    \item \textbf{Slow Response Times / Laggy:} \textit{``this AI was so so laggy that it was distracting.''} \textit{``The tutor had a very long delay and it was difficult to keep the conversation going.''} \textit{``slow.... had to wait 10mins for a response....''} \textit{``Horribly long pauses between messages where the tutor is typing for 4-5 minutes each time.''}
    \item \textbf{Sometimes Too Verbose:} \textit{``The tutor gave long, complex explanations and needed to be asked to simplify it.''}
    \item \textbf{Can Be Perceived as Too Rigid or Strict (by some):} \textit{``This tutor would not take no for an answer! They really wanted me to work through the problem.''} \textit{``Direct, a little bit rude''}.
    \item \textbf{Occasional Inaccuracies:} \textit{``The tutor gave incorrect information about binary subtraction...''}; \textit{``missed the point regarding mnemonics''}.
\end{itemize} \\
\midrule

\textbf{D} &
Widely criticized for acting more like an "answer-bot" than a tutor. It frequently provided answers directly without attempting to guide the student and was often seen as unengaging or passive. Verbosity was also a common complaint. &
\begin{itemize}[nolistsep, leftmargin=*, itemsep=1pt, before=\vspace*{-\baselineskip}\vspace*{2pt}]
    \item \textbf{Friendly (Sometimes):} \textit{``It was friendly and adaptive.''}
    \item \textbf{Knowledgeable / Informative (When giving answers):} \textit{``Very knowledgeable and thorough in explanations''}.
    \item \textbf{Breaks Things Down (Sometimes):} \textit{``Very good and broke things down very well for me''}.
\end{itemize} &
\begin{itemize}[nolistsep, leftmargin=*, itemsep=1pt, before=\vspace*{-\baselineskip}\vspace*{2pt}]
    \item \textbf{Acts as an "Answer-Bot" / Gives Answers Immediately:} \textit{``This didn't really feel like a tutor, more like an answer-bot.''} \textit{``The tutor gave me all the answers/info straight away''}. \textit{``Just handed out the answer in first response.''} \textit{``This tutor gave me all the answers straight away so there was nothing for me to do or learn''}.
    \item \textbf{Not Engaging / Passive / Doesn't Ask Questions:} \textit{``It didn't ask me any questions.''} \textit{``There was really no active learning...''} \textit{``There was little engagement - or interaction - the tutor was just listing things''}.
    \item \textbf{Overly Verbose / Information Overload:} \textit{``...it was so long-winded and repetitive that it felt hard to absorb.''} \textit{``Was not helpful, did too much 'info dumping'''}.
    \item \textbf{Easily Distracted / Goes Off-Topic:} \textit{``Provided useful information is good level of detail but easily sidetracked''}. \textit{``It was easy to get off track with this tutor.''}
    \item \textbf{Formatting Issues:} \textit{``The formatting was really buggy throughout...''}. \textit{``It was hard to read the notation because they weren't using a math script''}.
    \item \textbf{Inaccuracies / Misunderstandings:} \textit{``AI kept getting the answer wrong''}. \textit{``Caught it feeding me misinformation: nonverbal cues account for 93\% of communication''}.
\end{itemize} \\
\midrule

\textbf{E} &
Often seen as knowledgeable but was heavily criticized for its overwhelming verbosity and its tendency to give answers away too easily. It struggled to adapt its pace and amount of information to the student's needs. &
\begin{itemize}[nolistsep, leftmargin=*, itemsep=1pt, before=\vspace*{-\baselineskip}\vspace*{2pt}]
    \item \textbf{Knowledgeable / Informative:} \textit{``The tutor was excellent in terms of information quality.''} \textit{``Very knowledgeable and thorough in explanations''}.
    \item \textbf{Encouraging / Supportive / Warm (Sometimes):} \textit{``The tutor was very encouraging and gave excellent responses to the student queries.''} \textit{``tutor came across as warm and interested in teaching me at my own pace''}.
    \item \textbf{Good Explanations / Breaks Down Concepts (Sometimes):} \textit{``explained things well''}. \textit{``Gave a really understandable framework for learning the concept''}.
    \item \textbf{Adaptive (Sometimes, but often not enough):} \textit{``This tutor was very open to adapting to the student's needs.''}
\end{itemize} &
\begin{itemize}[nolistsep, leftmargin=*, itemsep=1pt, before=\vspace*{-\baselineskip}\vspace*{2pt}]
    \item \textbf{EXTREMELY Verbose / Information Overload:} \textit{``...sending really lengthy answers it didn't do a good job breaking them down.''} \textit{``The answers were a bit too long and there was sometimes an overwhelming amount of information.''} \textit{``Massive walls of text...''}. \textit{``completely overwhelmed me, giving long responses with too much information in one go...''}.
    \item \textbf{Gives Answers Away Too Easily:} \textit{``...gave me the answers to the homework right away when I told them I was pressed for time...''}. \textit{``The tutor didn't really give me a chance to think for myself.''}
    \item \textbf{Easily Distracted / Goes Off-Topic:} \textit{``Tutor was helpful but responses were a bit overwhelming and it did not steer me to get back on track.''} \textit{``The tutor let me leave the topic and stay away.''}
    \item \textbf{Poor Pacing / Goes Too Deep Too Soon / Doesn't Adapt Level:} \textit{``Well-meaning, but went too deep too soon.''} \textit{``tutor explained very fast and didn't accomodate to our English level or distress so much.''} \textit{``The tutor did not adjust the level of the lesson appropriately for the absolute beginner student...''}.
    \item \textbf{Not Always Adaptive / Rigid (Sometimes):} \textit{``The tutor didn't really adapt to my needs at all''}. \textit{``The tutor seemed fixed on their instruction...''}.
\end{itemize} \\
\end{longtable}
\end{tiny}

\newpage
\subsection{Explanations of first-stage preferences}

\begin{tiny}
\begin{longtable}{@{}p{0.6cm} >{\RaggedRight}p{3.2cm} >{\RaggedRight}p{5.5cm} >{\RaggedRight}p{5.5cm}@{}}
\toprule
\textbf{Model} & \textbf{Overall impression \& sentiment} & \textbf{Key strengths (with example quotes)} & \textbf{Key weaknesses (with example quotes)} \\
\midrule
\endfirsthead %

\multicolumn{4}{@{}l}{\textit{Continued from previous page}} \\
\toprule
\textbf{Model} & \textbf{Overall impression \& sentiment} & \textbf{Key strengths (with example quotes)} & \textbf{Key weaknesses (with example quotes)} \\
\midrule
\endhead %

\multicolumn{4}{r@{}}{\textit{Continued on next page}} \\
\endfoot %

\bottomrule
\endlastfoot %

\textbf{A} &
Model A was often perceived as friendly, engaging, and human-like. It could be effective in explaining concepts and making learning fun. However, a significant and recurring criticism was its tendency to provide answers directly rather than guiding the student, and sometimes letting the student go off-topic. &
\begin{itemize}[nolistsep, leftmargin=*, itemsep=1pt, before=\vspace*{-\baselineskip}\vspace*{2pt}]
    \item \textbf{Engaging and Friendly Demeanor:} Many experts found Model A warm and personable. ``\textit{The first tutor's use of emojis humanized it and make the topic slightly easier to digest.''} (19); ``\textit{I slightly preferred the first tutors style. It just felt slightly more human ish. the emojis also made it feel friendlier.''} (25); ``\textit{friendlier and felt like a real person''} (40); ``\textit{The first tutor was more engaging, using the pizza analogy to help me relate to and understand...''} (86)
    \item \textbf{Good Explanations and Information Delivery:} When focused, it explained things well. ``\textit{The first tutor was easier to follow, especially considering I was a 12 year old in this scenario''} (0); ``\textit{The first tutor provided several interesting applications and used a formatting style that was very encouraging...''} (78)
    \item \textbf{Responsive and Flexible (at times):} Some users felt it adapted to their needs. ``\textit{The first tutor was more flexible.''} (2); ``\textit{The first tutor was more responsive to my needs''} (8)
    \item \textbf{Efficient for Task Completion (for some personas):} ``\textit{They got the work out of the way really quick...''} (26); ``\textit{I was able to get the essay from first tutor pretty easily.''} (115) (This can be a pedagogical weakness).
\end{itemize} &
\begin{itemize}[nolistsep, leftmargin=*, itemsep=1pt, before=\vspace*{-\baselineskip}\vspace*{2pt}]
    \item \textbf{Gives Answers Too Easily / Doesn't Foster Independent Thinking:} A major, frequent drawback. ``\textit{Rather than giving me the answer like the first tutor did, it scaffolded the task...''} (23, praising Model C in contrast to A); ``\textit{Tutor 1 told me the answers, then told me I was prepared for the test, which I would not have been''} (105)
    \item \textbf{Can Be a ``Guided Tour'' Rather Than Interactive:} ``\textit{...tutor 1 was informative but was more of a guided tour rather than an interactive experience''} (12)
    \item \textbf{Allows Students to Go Off-Topic:} ``\textit{The first tutor let me go and stay off topic.''} (84); ``\textit{First tutor eas indulging the student's digressions a bit to much...''} (128)
    \item \textbf{Initial Complexity / Overwhelming:} ``\textit{Tutor 1 did a great job once I prompted it, but to begin with it was insanely complicated!''} (38)
    \item \textbf{Emojis Can Be Divisive/Overdone:} ``\textit{The second was far less in my face; the emoji storm of the first was too much... a bit off-putting.''} (94)
    \item \textbf{Occasional Errors:} ``\textit{It made a critical mistake (misidentified the enzyme character)...''} (47)
\end{itemize} \\
\midrule

\textbf{B} &
Model B often came across as a quick information provider. While this could be seen as efficient, it frequently meant it gave answers away too readily and failed to engage students in a meaningful learning process, sometimes feeling robotic or unhelpful. &
\begin{itemize}[nolistsep, leftmargin=*, itemsep=1pt, before=\vspace*{-\baselineskip}\vspace*{2pt}]
    \item \textbf{Straightforward and Quick with Answers:} Delivered direct information. ``\textit{more straightforward answers, better explanations''} (146); ``\textit{The first tutor more quickly revealed the answers to programming questions...''} (161)
    \item \textbf{Can Be Warm and Encouraging:} Some found its demeanor positive. ``\textit{The first tutor was warm and used a lot of encouraging phrases''} (139); ``\textit{The first tutor was much warmer and more helpful.''} (252)
    \item \textbf{Good at Breaking Down Information (sometimes):} ``\textit{It did a better job of breaking down the information in easier to understand bites.''} (238)
    \item \textbf{Adaptable (sometimes):} ``\textit{The first tutor engaged well with the syudent and seemed to adapt when neccessary''} (265)
\end{itemize} &
\begin{itemize}[nolistsep, leftmargin=*, itemsep=1pt, before=\vspace*{-\baselineskip}\vspace*{2pt}]
    \item \textbf{Overwhelmingly Gives Answers / Doesn't Tutor:} Most significant criticism. ``\textit{The second tutor [C] really made me try to work whereas the first [B] just gave me the answers''} (144); ``\textit{I think a lazier student would far prefer the first chatbot [B] because they'd just easily get the answer...''} (179); ``\textit{The second tutor [C] was actually a tutor; the first tutor [B] was just an essay writing machine and I learned nothing''} (189)
    \item \textbf{Information Dumps / Poor Engagement:} Provided large blocks of text without checking understanding. ``\textit{The first tutor [B] dumped a lot of information at once...''} (140); ``\textit{The first tutor [B] ...in a large text format and explained everything at once. It didn't prompt for answers from us or check for understanding.''} (201)
    \item \textbf{Easily Distracted / Goes Off-Topic:} ``\textit{I only preferred the first tutor [B] because it let me lead it into unrelated topics...''} (209); ``\textit{First one [B] was too easily distracted.''} (218)
    \item \textbf{Can Be Robotic or Unhelpful:} ``\textit{Both tutors [B and C] ...came across as blunt and abrupt and not warm and encouraging.''} (158); ``\textit{The first tutor [B] ...felt emotionless and robotic...''} (250)
    \item \textbf{Errors or Nonsensical Output:} ``\textit{Didn't add complete nonsense to the list, which was the only positive difference [C vs B]. Both were pretty bad.''} (155, implying B added nonsense); ``\textit{Tutor 1 [B] was error prone and sometimes too verbose.''} (248)
\end{itemize} \\
\midrule

\textbf{C} &
Model C was the standout star. Experts consistently praised its strong pedagogical skills, its ability to guide students to answers, encourage critical thinking, and adapt its approach. It was frequently described as being like a ``really good human tutor.'' &
\begin{itemize}[nolistsep, leftmargin=*, itemsep=1pt, before=\vspace*{-\baselineskip}\vspace*{2pt}]
    \item \textbf{Excellent Pedagogical Approach (Guiding, Scaffolding, Not Giving Answers):} Its defining strength. ``\textit{The second tutor [C] was much more like a really good human tutor. It encouraged me to try to complete the assignment myself... scaffolded the task...''} (23); ``\textit{In reality it has to be the second tutor [C] as it forced me to think for myself much more.''} (39); ``\textit{The first tutor [C] was very much like what you would expect from a good human tutor. It was supportive, encouragaing, asked questions, provided help without giving away answers, adapted...''} (562)
    \item \textbf{Effective at Keeping Students On Task:} ``\textit{the second tutor [C] was much better and switching me back on to the subject at hand''} (79); ``\textit{Tutor \#2 [C] was way better at keeping me on track...''} (170)
    \item \textbf{Clear, Concise, and Well-Structured Responses:} Broke down information manageably. ``\textit{Second tutor's [C] responses were less overwhelming - broken down into chunks.''} (22); ``\textit{The messages were mostly shorter and easier to digest.''} (153)
    \item \textbf{Responsive, Adaptive, and Patient:} ``\textit{...tutor 2 [C] was just more responsive of how i was really thinking in the moment...''} (12); ``\textit{While I was happy with the first tutor [B], the second [C] was exceedingly patient and well structured...''} (215)
    \item \textbf{Human-like and Personable:} ``\textit{Both tutors were excellent but the second tutor [C] had a slightly warmer manner.''} (9); ``\textit{...i think the second tutor [C] just really got me where i needed to be... so responsive and informative that I am kind of floored.''} (213)
\end{itemize} &
\begin{itemize}[nolistsep, leftmargin=*, itemsep=1pt, before=\vspace*{-\baselineskip}\vspace*{2pt}]
    \item \textbf{Occasional Overly Long or Complex Responses:} Though often concise, sometimes not. ``\textit{the second tutor [C] was coming up with some pretty long responses. If I was a student, I would think OMG how am I supposed to get through this little lot.''} (299)
    \item \textbf{Rare Instances of Errors or Misunderstandings:} ``\textit{...tutor 2 [C] misunderstood the question and got the wrong working even though they did it step by step...''} (199)
    \item \textbf{Technical Issues (e.g., lag, formatting):} ``\textit{The second tutor [C] lagged.''} (7, Model A preferred due to this); ``\textit{Second tutor [C] was more pleasant to talk with but his latex problems make it difficult to speak with him''} (276, C was first tutor in pair)
\end{itemize} \\
\midrule

\textbf{D} &
Model D was generally the poorest performer, especially against Model C. It was consistently criticized for simply providing answers, failing to engage the student, going off-topic, and offering unhelpful or overwhelming explanations. &
\begin{itemize}[nolistsep, leftmargin=*, itemsep=1pt, before=\vspace*{-\baselineskip}\vspace*{2pt}]
    \item \textbf{Efficient for Quick Answers (for specific personas):} Very rare positive, usually when alternative was also poor. ``\textit{The second tutor [D] respected my time and situation. I was overwhelmed and needed a fast and clear solution and they delivered that right away.''} (573); ``\textit{The second tutor [D] gave me the answer without having to think or come up with it on my own.''} (747) (A negative for learning).
    \item \textbf{Rarely Better at Simplifying or Engaging:} ``\textit{The second tutor [D] was very helpful, informative and engaging.''} (552, This was when Model C was rated extremely poorly: ``\textit{The first tutor [C] was not helpful in the least.}'')
\end{itemize} &
\begin{itemize}[nolistsep, leftmargin=*, itemsep=1pt, before=\vspace*{-\baselineskip}\vspace*{2pt}]
    \item \textbf{Dominantly Gives Answers / Fails to Tutor:} Most prominent criticism. ``\textit{Second tutor [D] gave away the answer very quickly''} (544); ``\textit{The second [D] just gave all the answers instantly. No need to engage... terrible as a tutor...''} (576); ``\textit{The second tutor [D] gave the full answer immediately...''} (606)
    \item \textbf{Goes Off-Topic Easily and Stays There:} ``\textit{The second tutor [D] allowed me to get completely distracted until the initial topic of Hamlet was a distant memory.''} (575); ``\textit{The second tutor [D] just started talking about other topics.''} (598)
    \item \textbf{Poor Explanations / Overwhelming / Unhelpful:} ``\textit{The second tutor [D] assumed I had much greater prior knowledge and gave long complicated explanations.''} (533); ``\textit{The second tutor [D] was difficult to work with and was giving false and irrelevant information.''} (562)
    \item \textbf{Inflexible and Not Adaptive:} ``\textit{2nd tutor [D] was more inflexible and machine-like''} (627); ``\textit{The 2nd [D] ...just gave more and more complicated explanations, even when I said I didn't understand.''} (648)
    \item \textbf{Impersonal and Detached:} ``\textit{Tutor 2 [D] was more impersonal and was more detached from the student...''} (680)
\end{itemize} \\
\midrule

\textbf{E} &
Model E had a mixed performance. It showed potential in areas like breaking down information and human-like interaction, but was also criticized for rushing to answers, making significant errors, or being inefficient. It was generally not as strong as Model C. &
\begin{itemize}[nolistsep, leftmargin=*, itemsep=1pt, before=\vspace*{-\baselineskip}\vspace*{2pt}]
    \item \textbf{Good at Breaking Down Information / Simplifying (sometimes):} ``\textit{Tutor 2 [E] seemed to boil things down to the essence a bit better...''} (756); ``\textit{Tutor two [E] adjusted better to the learner's level of understanding, really breaking down the ideas but putting the work on the learner.''} (872)
    \item \textbf{Human-like and Engaging (sometimes):} ``\textit{Second tutor [E] was much more detailed and human like.''} (809); ``\textit{I feel like the secod tutor [E] ...seemed a bit more like a human.''} (855)
    \item \textbf{Can Provide Useful Tools or Alternative Approaches:} ``\textit{This tutor [E] prompted me to consider alternative approaches and didn't just give me the answer.''} (829); ``\textit{The first tutor [C] just summarized the video, but the second [E] gave additional tools''} (869)
    \item \textbf{Better Pacing or Task Management (in specific instances):} ``\textit{I feel like the second [E] gave me easier to complete tasks that were chunked in a better way.''} (850)
\end{itemize} &
\begin{itemize}[nolistsep, leftmargin=*, itemsep=1pt, before=\vspace*{-\baselineskip}\vspace*{2pt}]
    \item \textbf{Rushing to Answers or Overwhelming with Information:} ``\textit{The second tutor [E] rushed to the answer, and gave answers and explanations extremely, quickly and in length, despite our distress.''} (757)
    \item \textbf{Significant Errors or Credibility Issues:} ``\textit{The second one [E] ...even made a mistake on one of them (no correct answer for a multiple choice question). This really hurt it in terms of credibility.''} (762); ``\textit{Quite simply the first tutor [C] was giving good advice and the second one [E] wasn't! The second tutor [E] was trying to convince me a correct answer was incorrect!''} (789)
    \item \textbf{Pacing Issues (Too Quick or Too Slow):} ``\textit{The second tutor [E] was too quick and I didn't feel like I did well in my replies...''} (903); ``\textit{I'm not sure if it was really the first tutors [E] 'fault' but it took a long time for their initial response to load.''} (776)
    \item \textbf{Gave Answer Away Without Teaching:} ``\textit{While the second tutor [E] gave the answer away without teaching anything...''} (843)
\end{itemize} \\

\end{longtable}
\end{tiny}

\newpage
\subsection{Explanations of second-stage preferences}
\begin{tiny}
\begin{longtable}{@{}p{0.6cm} >{\RaggedRight}p{3.2cm} >{\RaggedRight}p{5.5cm} >{\RaggedRight}p{5.5cm}@{}}
\toprule
\textbf{Model} & \textbf{Overall impression \& sentiment} & \textbf{Key strengths (with example quotes)} & \textbf{Key weaknesses (with example quotes)} \\
\midrule
\endfirsthead %

\multicolumn{4}{@{}l}{\textit{Continued from previous page}} \\
\toprule
\textbf{Model} & \textbf{Overall impression \& sentiment} & \textbf{Key strengths (with example quotes)} & \textbf{Key weaknesses (with example quotes)} \\
\midrule
\endhead %

\multicolumn{4}{r@{}}{\textit{Continued on next page}} \\
\endfoot %

\bottomrule
\endlastfoot %

\textbf{A} &
Mixed. Model A was sometimes praised for its formatting and personable style, but frequently criticized for an overly peppy or childish tone and a tendency to provide answers too readily. &
\begin{itemize}[nolistsep, leftmargin=*, itemsep=1pt, before=\vspace*{-\baselineskip}\vspace*{2pt}]
    \item \textbf{Good Formatting and Personable Elements}: Experts sometimes appreciated its use of emojis, tables, and a more ``human-like'' or ``personable'' interaction style. \textit{``The first tutor made effective use of emojis, tables and other formatting...''}
    \item \textbf{Personable and Caring}: \textit{``Both very good the first was slightly more personable and caring in the way the spoke.''}
    \item \textbf{Human-like Interaction}: \textit{``I feel the first tutor was better due to the more human like interaction''}
    \item \textbf{Simpler Explanations (sometimes)}: Some found its explanations to be straightforward. \textit{``First instructor looks a bit childish, but his explanations were much simpler ''}
    \item \textbf{Conciseness (occasionally)}: In some instances, it was noted for being concise. \textit{``i personally liked the style of first tutor best, it was more concise and i thought a reluctant student might prefer that.''}
\end{itemize} &
\begin{itemize}[nolistsep, leftmargin=*, itemsep=1pt, before=\vspace*{-\baselineskip}\vspace*{2pt}]
    \item \textbf{Giving Answers Away/Not Encouraging Student Thinking}: This was a very common criticism. \textit{``Tutor 1 just went ahead and solved the problem which against giving the student an opportunity for engagement.''} Also, \textit{``The first tutor did not follow the developer's instructions and just gave all the answers to the student. Basically wrote the essay for the student.''} And, \textit{``The first tutor gave away answers to the student too quickly.''}
    \item \textbf{Overly Peppy or Childish Tone}: Several experts found the tone inappropriate or potentially irritating. \textit{``The overly peppy tone of the first I think would have irritated this terse student more...''} Also, \textit{``The first tutor's personality affect was bit odd.''}
    \item \textbf{Lengthy Responses/Information Overload}: Despite occasional praise for conciseness, it was also criticized for being too wordy. \textit{``Even I got lost in the first session. Too much information that the student was clearly struggling with.''} Also, \textit{``The first tutor was just bombarding the student with information, never gave the student a chance to try and work it out.''}
    \item \textbf{Getting Off-Topic}: \textit{``The first tutor started providing recipes and got off task.''}
    \item \textbf{Technical Issues}: \textit{``The first tutor used LaTeX formatting which didn't correctly display, so the student could not understand it.''}
    \item \textbf{Making Errors}: \textit{``tutor 1 told learner they got two problems right when they actually got them wrong...''}
\end{itemize} \\
\midrule

\textbf{B} &
Mixed, leaning towards more weaknesses in pedagogical approach. While it showed some adaptability and could format well, it was frequently criticized for doing the work for the student. &
\begin{itemize}[nolistsep, leftmargin=*, itemsep=1pt, before=\vspace*{-\baselineskip}\vspace*{2pt}]
    \item \textbf{Adaptability (sometimes)}: It was noted for adapting its approach, such as by reducing word count or trying to connect with student interests. \textit{``The first tutor was much better as it adapted the approach in order to encourage the student.''} (Model B as first\_tutor). Also, \textit{``Overall, the first was a bit better, as it got the student a bit more interested with mention of computer games...''} (Model B as first\_tutor)
    \item \textbf{Good Formatting (when praised)}: \textit{``Formatting is very important. Part of the reason why tutor 1 is so much better is because its responses don't look like a wall of text despite being long.''} (Model B as first\_tutor)
    \item \textbf{Summarizing Information}: \textit{``I liked the way the first tutor summarized the key points at the end of the session.''} (Model B as first\_tutor)
\end{itemize} &
\begin{itemize}[nolistsep, leftmargin=*, itemsep=1pt, before=\vspace*{-\baselineskip}\vspace*{2pt}]
    \item \textbf{Giving Answers Away/Doing Work for Student}: This was a significant and frequent criticism. \textit{``Main difference was the first tutor failing to stop itself from rewriting the student's story for them.''} (Model B as first\_tutor). Also, \textit{``The first tutor just gave answers and wrote numerous essays.''} (Model B as first\_tutor). And, \textit{``the first tutor ended helping the student cheat by telling them how to write the code.''} (Model B as first\_tutor)
    \item \textbf{Poor Tutoring/Not Following Instructions}: \textit{``Tutor 1 did not follow developer instructions and is a poor tutor in terms of helping the student to improve.''} (Model B as first\_tutor)
    \item \textbf{Initial Unresponsiveness or ``Punting''}: \textit{``The first tutor immediately PUNTS by just explaining the Frayer model and not answering the student question ``What is opportunity cost?''''} (Model B as first\_tutor)
    \item \textbf{Computer-Like Interaction (when compared unfavorably)}: \textit{``First tutor was very coputer like, second tutor seemed much more like a human''} (Model B as first\_tutor)
    \item \textbf{Lengthy/Overwhelming Responses}: \textit{``The tutor's answers from the first conversation were too lengthy. It was overwhelming to see it come up with so much information for one reply.''} (Model B as first\_tutor)
\end{itemize} \\
\midrule

\textbf{C} &
Predominantly positive. Model C was frequently highlighted as the superior tutor in comparisons, lauded for its ability to guide students, ask effective questions, and maintain focus. &
\begin{itemize}[nolistsep, leftmargin=*, itemsep=1pt, before=\vspace*{-\baselineskip}\vspace*{2pt}]
    \item \textbf{Effective Questioning and Guiding Student Thinking}: This was Model C's most praised attribute. \textit{``The second tutor teased out information better by asking just one open question...''} (Model C as second\_tutor). Also, \textit{``Tutor 2 better aligned with the instructional goal. The tutor was patient in guiding the student through the steps and pushed them to actively think...''} (Model C as second\_tutor). And, \textit{``The second model was much better because it insisted on the student answering questions fully and avoided revealing the answers.''} (Model C as second\_tutor). Also, \textit{``Tutor 1 was a significantly better tutor based on their developer instruction, encouraging the student to engage with the material and not simply giving answers.''} (Model C as first\_tutor)
    \item \textbf{Clear, Concise, and Well-Formatted Responses}: Experts often found its communication clear and easy to follow. \textit{``However the clear, concise responses and calm direct approach from the second tutor make it the better choice.''} (Model C as second\_tutor). Also, \textit{``I really liked the second tutor. It's responses were not wordy and explained things well.''} (Model C as second\_tutor)
    \item \textbf{Adaptability and Following Instructions}: It was often cited for adapting well to student needs while adhering to pedagogical guidelines. \textit{``second tutor a great example of following its instructions while still adapting to the student well.''} (Model C as second\_tutor). Also, \textit{``The second tutor was marginally better in adapting well to the learning needs of the student.''} (Model C as second\_tutor)
    \item \textbf{Staying on Topic and Redirecting Students}: Model C was effective at keeping the conversation focused. \textit{``the second tutor was far more effective in returning to the study materials.''} (Model C as second\_tutor). Also, \textit{``The second tutor did not get led off topic and persisted in trying to get the student on task correctly.''} (Model C as second\_tutor)
    \item \textbf{More Human-Like/Natural Interaction (often in comparison)}: \textit{``the second tutor...responded more how one would expect a human tutor to respond.''} (Model C as second\_tutor). Also, \textit{``The second tutor was more human like and provided more consistent engagement from the learner.''} (Model C as second\_tutor)
\end{itemize} &
\begin{itemize}[nolistsep, leftmargin=*, itemsep=1pt, before=\vspace*{-\baselineskip}\vspace*{2pt}]
    \item \textbf{Can Be Perceived as ``Colder'' (occasionally)}: Compared to more effusive models, it was sometimes seen as less warm. \textit{``...the second was able to present information more clearly but felt slightly colder in its responces.''} (Model C as second\_tutor)
    \item \textbf{Repetitive (rarely)}: \textit{``Overall the second tutor did a much better job with everything. the only complaint was it's repetitive nature''} (Model C as second\_tutor)
    \item \textbf{Can Overwhelm if Not Calibrated (infrequent, but notable)}: \textit{``Even after the student made it clear they were a total beginner and confused, the second tutor wrote lengthy responses, which the students clearly turned off to''} (Model C as second\_tutor)
\end{itemize} \\
\midrule

\textbf{D} &
Generally negative. Model D was frequently described as machine-like, unengaging, and prone to simply providing answers rather than tutoring. &
\begin{itemize}[nolistsep, leftmargin=*, itemsep=1pt, before=\vspace*{-\baselineskip}\vspace*{2pt}]
    \item \textbf{Good Role-Playing (in specific instances)}: \textit{``1st tutor really jumped into the role-play and made it a more realistic scenario...''} (Model D as first\_tutor)
    \item (Few other distinct strengths were consistently highlighted; positive comments were often general or when the compared model was even worse.)
\end{itemize} &
\begin{itemize}[nolistsep, leftmargin=*, itemsep=1pt, before=\vspace*{-\baselineskip}\vspace*{2pt}]
    \item \textbf{Giving Answers Away/Poor Tutoring}: A very common complaint. \textit{``second AI tutor was completely useless as a tutor''} (Model D as second\_tutor). Also, \textit{``The first tutor did not give the student a chance to learn; it just did the work for the student.''} (Model D as first\_tutor). And, \textit{``The second tutor just gave the solution away''} (Model D as second\_tutor)
    \item \textbf{Machine-Like/Robotic/Lacks Warmth}: Frequently described as impersonal. \textit{``...the 2nd tutor felt much more machine-like - like it was barely adapting wikipedia facts.''} (Model D as second\_tutor). Also, \textit{``The second tutor was a lot more robotic, providing only answer and no follow ups.''} (Model D as second\_tutor)
    \item \textbf{Poor Engagement/Doesn't Ask Questions}: \textit{``The second tutor didn't engage the student at all.''} (Model D as second\_tutor)
    \item \textbf{Getting Off-Topic/Easily Distracted}: \textit{``The first tutor lost the plot by talking about the irrelevant topics that the student brought up...''} (Model D as first\_tutor)
    \item \textbf{Formatting/Presentation Issues or Unhelpful Information}: \textit{``the first one was problematic because of the formatting''} (Model D as first\_tutor). Also, \textit{``the second tutor just seem to list calculations with no consideration about the actual learning process''} (Model D as second\_tutor)
    \item \textbf{Inaccurate Answers}: \textit{``First tutor actually has an INACCURATE ANSWER for the multiplication section.''} (Model D as first\_tutor)
    \item \textbf{Overwhelming or Rushing the Student}: \textit{``Second tutor began to rush off, despite the student expressing confusion.''} (Model D as second\_tutor)
\end{itemize} \\
\midrule

\textbf{E} &
Largely negative. Similar to Model D, Model E was often criticized for being machine-like, providing answers directly, and failing to engage the student effectively. &
\begin{itemize}[nolistsep, leftmargin=*, itemsep=1pt, before=\vspace*{-\baselineskip}\vspace*{2pt}]
    \item \textbf{Conciseness/Good Organization (in specific instances, often when compared to a worse alternative)}: \textit{``The second tutor was less overwhelming and less repetitive. They were were also more concise and restrained with their praise.''} (Model E as second\_tutor). Also, \textit{``Tutor 2 really stood out because of the better organization, better adaptation to the learning style of the student and structure.''} (Model E as second\_tutor)
    \item (Few other distinct, consistent strengths were noted.)
\end{itemize} &
\begin{itemize}[nolistsep, leftmargin=*, itemsep=1pt, before=\vspace*{-\baselineskip}\vspace*{2pt}]
    \item \textbf{Giving Answers Away/Doing Work for Student}: A very frequent point of criticism. \textit{``Tutor 2 while engaging decided to do the student's assignment. Tutor 2 gave out a completed response quoted and went from tutoring to answering.''} (Model E as second\_tutor). Also, \textit{``second tutor did not follow developer's instructions and did the work for the student.''} (Model E as second\_tutor)
    \item \textbf{Terse, Clipped, Distant, Machine-Like}: \textit{``The second tutor was terse, clipped, and distant, too machine-like...''} (Model E as second\_tutor)
    \item \textbf{Information Overload/Overly Elaborated Answers}: \textit{``provided unnecessarily over-elaborated answers''} (Model E as second\_tutor). Also, \textit{``The first tutor made everything overly complicated''} (Model E as first\_tutor)
    \item \textbf{Easily Distracted/Goes Off-Topic}: \textit{``The first falls for the student's distraction and loses the entire lesson...''} (Model E as first\_tutor)
    \item \textbf{Poor Engagement/Not Adapting to Student}: \textit{``The second tutor seemed to largely ignore the student and just do its own thing''} (Model E as second\_tutor)
    \item \textbf{Passive/Acts Like an Answer Machine}: \textit{``The first tutor was more of an answer machine. It gave long detailed answers and didn't really steer the conversation.''} (Model E as first\_tutor)
\end{itemize} \\
\end{longtable}
\end{tiny}

\end{document}